\documentclass[twocolumn, twocolappendix]{aastex631}

\usepackage{chngcntr}

\begin{document}

\title{EPOCHS XI: The Structure and Morphology of Galaxies in the Epoch of Reionization to $z \sim 12.5$}

\author[0009-0008-8642-5275]{Lewi Westcott}
\affiliation{Jodrell Bank Centre for Astrophysics, Alan Turing Building, University of Manchester, Oxford Road, Manchester M13 9PL, UK}

\author[0000-0003-1949-7638]{Christopher J. Conselice}
\affiliation{Jodrell Bank Centre for Astrophysics, Alan Turing Building, University of Manchester, Oxford Road, Manchester M13 9PL, UK}

\author[0000-0002-4130-636X]{Thomas Harvey}
\affiliation{Jodrell Bank Centre for Astrophysics, Alan Turing Building, University of Manchester, Oxford Road, Manchester M13 9PL, UK}

\author[0000-0003-0519-9445]{Duncan Austin}
\affiliation{Jodrell Bank Centre for Astrophysics, Alan Turing Building, University of Manchester, Oxford Road, Manchester M13 9PL, UK}

\author[0000-0003-4875-6272]{Nathan Adams}
\affiliation{Jodrell Bank Centre for Astrophysics, Alan Turing Building, University of Manchester, Oxford Road, Manchester M13 9PL, UK}

\author[0000-0002-0056-1970]{Fabricio Ferrari}
\affiliation{Instituto de Matemática Estatística e Física, Universidade Federal do Rio Grande, Rio Grande, RS, Brazil}

\author[0000-0002-8919-079X]{Leonardo Ferreira}
\affiliation{Department of Physics \& Astronomy, University of Victoria, Finnerty Road, Victoria, British Columbia, V8P 1A1, Canada}
\author[0000-0002-9081-2111]{James Trussler}
\affiliation{Jodrell Bank Centre for Astrophysics, Alan Turing Building, University of Manchester, Oxford Road, Manchester M13 9PL, UK}

\author[0000-0002-3119-9003]{Qiong Li}
\affiliation{Jodrell Bank Centre for Astrophysics, Alan Turing Building, University of Manchester, Oxford Road, Manchester M13 9PL, UK}

\author[0000-0001-7633-3985]{Vadim Rusakov}
\affiliation{Jodrell Bank Centre for Astrophysics, Alan Turing Building, University of Manchester, Oxford Road, Manchester M13 9PL, UK}

\author[0009-0009-8105-4564]{Qiao Duan}
\affiliation{Jodrell Bank Centre for Astrophysics, Alan Turing Building, University of Manchester, Oxford Road, Manchester M13 9PL, UK}

\author[0009-0005-0817-6419]{Honor Harris}
\affiliation{Jodrell Bank Centre for Astrophysics, Alan Turing Building, University of Manchester, Oxford Road, Manchester M13 9PL, UK}

\author[0009-0006-4799-4497]{Caio Goolsby}
\affiliation{Jodrell Bank Centre for Astrophysics, Alan Turing Building, University of Manchester, Oxford Road, Manchester M13 9PL, UK}

\author[0000-0002-8785-8979]{Thomas J. Broadhurst}
\affiliation{Department of Physics, University of the Basque Country UPV/EHU, E-48080 Bilbao, Spain}
\affiliation{DIPC, Basque Country UPV/EHU, E-48080 San Sebastian, Spain}
\affiliation{Ikerbasque, Basque Foundation for Science, E-48011 Bilbao, Spain}

\author[0000-0001-7410-7669]{Dan Coe} 
\affiliation{Space Telescope Science Institute, 3700 San Martin Drive, Baltimore, MD 21218, USA}
\affiliation{Association of Universities for Research in Astronomy (AURA) for the European Space Agency (ESA), STScI, Baltimore, MD 21218, USA}
\affiliation{Center for Astrophysical Sciences, Department of Physics and Astronomy, The Johns Hopkins University, 3400 N Charles St. Baltimore, MD 21218, USA}

\author[0000-0003-3329-1337]{Seth H. Cohen} 
\affiliation{School of Earth and Space Exploration, Arizona State University,
Tempe, AZ 85287-1404, USA}

\author[0000-0001-9491-7327]{Simon P. Driver} 
\affiliation{International Centre for Radio Astronomy Research (ICRAR) and the
International Space Centre (ISC), The University of Western Australia, M468,
35 Stirling Highway, Crawley, WA 6009, Australia}

\author[0000-0002-9816-1931]{Jordan C. J. D'Silva} 
\affiliation{International Centre for Radio Astronomy Research (ICRAR) and the
International Space Centre (ISC), The University of Western Australia, M468,
35 Stirling Highway, Crawley, WA 6009, Australia}
\affiliation{ARC Centre of Excellence for All Sky Astrophysics in 3 Dimensions
(ASTRO 3D), Australia}

\author[0000-0003-1625-8009]{Brenda Frye} 
\affiliation{Department of Astronomy/Steward Observatory, University of Arizona, 933 N Cherry Ave,
Tucson, AZ, 85721-0009, USA}

\author[0000-0001-9440-8872]{Norman A. Grogin} 
\affiliation{Space Telescope Science Institute,
3700 San Martin Drive, Baltimore, MD 21218, USA}

\author[0000-0001-6145-5090]{Nimish P. Hathi}
\affiliation{Space Telescope Science Institute, 3700 San Martin Drive, Baltimore, MD 21218, USA}

\author[0000-0003-1268-5230]{Rolf A. Jansen} 
\affiliation{School of Earth and Space Exploration, Arizona State University,
Tempe, AZ 85287-1404, USA}

\author[0000-0002-6610-2048]{Anton M. Koekemoer} 
\affiliation{Space Telescope Science Institute,
3700 San Martin Drive, Baltimore, MD 21218, USA}

\author[0000-0001-6434-7845]{Madeline A. Marshall} 
\affiliation{Los Alamos National Laboratory, Los Alamos, NM 87545, USA}

\author[0000-0002-6150-833X]{Rafael {Ortiz~III}} 
\affiliation{School of Earth and Space Exploration, Arizona State University,
Tempe, AZ 85287-1404, USA}

\author[0000-0003-3382-5941]{Nor Pirzkal} 
\affiliation{Space Telescope Science Institute,
3700 San Martin Drive, Baltimore, MD 21218, USA}

\author[0000-0003-0429-3579]{Aaron Robotham} 
\affiliation{International Centre for Radio Astronomy Research (ICRAR) and the
International Space Centre (ISC), The University of Western Australia, M468,
35 Stirling Highway, Crawley, WA 6009, Australia}

\author[0000-0003-0894-1588]{Russell E. Ryan, Jr.} 
\affiliation{Space Telescope Science Institute,
3700 San Martin Drive, Baltimore, MD 21218, USA}

\author[0000-0002-7265-7920]{Jake Summers} 
\affiliation{School of Earth and Space Exploration, Arizona State University,
Tempe, AZ 85287-1404, USA}

\author[0000-0001-9262-9997]{Christopher N. A. Willmer} 
\affiliation{Steward Observatory, University of Arizona,
933 N Cherry Ave, Tucson, AZ, 85721-0009, USA}

\author[0000-0001-8156-6281]{Rogier A. Windhorst}
\affiliation{School of Earth and Space Exploration, Arizona State University,
Tempe, AZ 85287-1404, USA}

\author[0000-0001-7592-7714]{Haojing Yan} 
\affiliation{Department of Physics and Astronomy, University of Missouri,
Columbia, MO 65211, USA}



\begin{abstract}

We present a structural analysis of 520 galaxy candidates at $6.5 < z < 12.5$, with SNR $> 10\sigma$ in the F444W filter, taken from the EPOCHS v1 sample, consisting of uniformly reduced deep JWST NIRCam data, covering the CEERS, JADES GOOD-S, NGDEEP, SMACS0723, GLASS and PEARLS surveys. We use standard software to fit single S\'ersic models to each galaxy in the rest-frame optical and extract their parametric structural parameters (S\'ersic index, half-light radius and axis-ratio), and \texttt{Morfometryka} to measure their non-parametric concentration and asymmetry parameters. We find a wide range of sizes for these early galaxies, with galaxy sizes overall continuing to become progressively smaller in the high-redshift regime, following $R_{e} = \left( 2.12 \pm 0.28 \right) \left( 1 + z \right) ^{-0.67 \pm 0.06}$ kpc. We further find a galaxy size-mass correlation up to $z \sim 12$, with galaxies of a given mass also becoming smaller. Using non-parametric methods we find that galaxy merger fractions, classified through asymmetry parameters, at these redshifts remain consistent with those in literature, maintaining a value of $f_{m} \sim 0.12 \pm 0.07$ showing little dependence with redshift when combined with literature at $z > 4$. We find that galaxies which are smaller in size also appear rounder, with an excess of high axis-ratio objects. Finally, we artificially redshift a subsample of our objects to determine how robust the observational trends we see are, determining that observed trends are due to real evolutionary effects, rather than being a consequence of redshift effects.

\end{abstract}



\section{Introduction} 
\label{sec:intro}

Since the launch of the James Webb Space Telescope (JWST), our capability to pursue one of the oldest studied properties in galaxies -  structure -  has seen an immense advancement. The structural nature of galaxies has long been of interest, dating back to when these objects beyond our own galaxy were first observed and commented on \citep{Herschel_1786} in the 18th century. It was clear then that these systems differed to the objects we see in our own galaxy, and that their structures hold clues to their formation. These structures and resolved properties are in many ways one of the oldest studied properties of galaxies \citep[e.g.][]{Herschel_1786, Buitrago2008, vanderWel_2012, Conselice_2014, Kartaltepe_2023, Seuss_2022, Ferreira_2022_disk_panic}.

As the studies of galaxies unfolded, our understanding and interest in these objects has deepened. From observing the spiral arm structure of galaxies for the first time \citep{Rosse_1850}, to the work of \citet{1926ApJ....64..321H}, with the emergence of photography, developing a system to differentiate classifications of galaxies with different visual appearances, now known as the `Hubble Tuning Fork' in which galaxies are split between `late-type' spiral galaxies, both with and without bars, `early-type' galaxies and irregular galaxies.

Further developments in structural analysis of galaxies, allowed for quantitative analysis of galaxies, with \citet{de_Vaucouleurs_1948} identifying that all bright ellipticals follow roughly the same light profile, now known as the de Vaucouleurs profile. This was later generalised, after it was found that all galaxies follow a more general form for light profiles, defined by the S\'ersic index (n) \citep{1963BAAA....6...41S}.

Developments have continued right up to the present with telescope observations, first through the Hubble Space Telescope (HST) and now with JWST, giving us the capabilities to observe galaxies far beyond the scope of our local universe, enabling us to observe the first galaxies from within the Epoch of Reionisation. Over the past 30 years, HST has pioneered the observations of galaxy structure in the rest frame optical up to $z \sim 3$, with HST observations showing us that galaxies appear to become progressively smaller, more irregular and more peculiar as redshift increases \citep[e.g.][]{Conselice_2003, Lotz_2004, Delgado_Serrano_2010, Mortlock_2013, Buitrago_2014, Conselice_2014, Schawinski_2014, Whitney_2021}.

While HST probed further than ever before, significant limitations prevented observations in the rest-frame optical, with the reddest filter on HST, F160W on Hubble's Wide Field Camera 3 (WFC3), only probing the rest-frame optical out to $z \sim 2.8$. We know that there are many galaxies which exist beyond this \citep[e.g.][]{Castellano_2022, Donnan_2022, Naidu_2022, Adams_2023, Atek_2023, Austin_2023, Curtis-Lake_2023, Finkelstein_2023, Harikane_2023, Yan_2023, Adams_2024}. The launch of the James Webb Space Telescope now allows us to probe to much higher redshifts than what HST was capable of. On board the telescope, the Near Infrared Camera (NIRCam, \citet{Rieke_2005, Rieke_2023b}) allows us to observe galaxies in the rest-frame optical out to $z \sim 9$, with filters as red as $4.8 \mu m$ (F480M). Beyond this, JWST also boasts a superior resolution to better spatially resolve structural features, giving us much more fidelity than what was previously possible.

Since the launch of JWST, we have seen a great diversity in morphology in the early universe. The earliest JWST structural observations proved the benefit of the instrument's superior fidelity, with many galaxies being re-classified from having a peculiar structure to being more disc-dominated, with spiral features being directly observed out to $z > 3$ \citep{Ferreira_2022_disk_panic, Ferreira_2023, Jacobs_2023, Nelson_2023, Robertson_2023, Kuhn_2024, xiao_2024}. As evidenced through their morphologies, we have further seen the different evolutionary paths star forming galaxies (SFGs) take to quiescent galaxies (QGs) \citep{Benton_2024, Clausen_2024, Martorano_2024, Ward_2024}, and how galaxy mergers play a major role in the increasing size and mass of galaxies at the highest redshifts \citep{dalmasso2024ratecontributionmergersmass, duan_2024}

Structural analysis at $z > 6$ however, is still a relatively understudied topic, and one where further analysis is needed to better uncover how the first galaxies formed and how they evolve to what we see in our local universe.   Even the very basic measurements of galaxy structure in the early universe are lacking, with an immense amount of work left to be done to quantify and measure and interpret early galaxy structure to infer evolution. 

In this paper, we present a comprehensive quantitative analysis of the structures of galaxies within the early universe, from $z > 6.5$, making use of the EPOCHS v1 catalogue \citep{Conselice_2024}, utilising both parametric and non-parametric structural measurements. We measure galaxy sizes and shapes based on S\'ersic profile fitting, to determine how structure evolves from the early universe, comparing to prior work at lower redshifts, $z < 8$. Further to this, we measure the concentration (C) and Asymmetry (A) non-parametric parameters, which we use to determine the fraction of mergers we see in the early universe. The evolution of these properties give us a key insight into the evolution of galaxies, and trace the processes which drive galaxy growth and quenching processes.

The structure of this paper is as follows: Section \ref{sec:data} discusses the data we use in this study and the data reduction process. Section \ref{sec:Catalogue selection} discusses our selection process for our high-redshift catalogue. We explain our fitting process in Section \ref{sec:morphological fitting} as well as our selection criteria based on these fits in Section \ref{sec:sample selection}. In Section \ref{sec:results} we present our results, along with a discussion of our results in \ref{sec:discussion}, and a summary of the paper in \ref{sec:conclusions}.

We assume a standard $\Lambda$CDM cosmology throughout, of $\Omega_{m} = 0.3$, $\Omega_{\Lambda} = 0.7$ and $H_{0} = 70\ {\rm kms^{-1}\ Mpc^{-1}}$. Where we refer to a galaxies `size', we are referring to the half-light radius ($R_{e}$) of the object. All magnitudes are given in the AB system throughout \citep{Oke_1974, Oke_&_Gunn_1983}.


\section{Data and processing} 
\label{sec:data}

In this work we use the EPOCHS survey, a catalog consisting of 1165 high redshift $6.5 < z < 18$ galaxies, found within a combination of JWST fields, covering more than 200 arcmin$^{2}$. The fields used are from both JWST public fields, and fields from the \textit{JWST Prime Extragalactic Areas for Reionization and Lensing Science} (PEARLS, PI: R. Windhorst \& H. Hammel, PID: 1176 \& 2738) GTO survey \citep{windhorst_2023}. Each field has been reduced homogeneously, as to give consistent image and data quality for the catalog as a whole. Here we give a brief description of each field used and the reduction process. Further details of the fields used can be found in \citet{Austin_2023, Adams_2024, Harvey2024}, whilst further details of the EPOCHS catalog can be found in \citet{Conselice_2024}. In this study we use only the F444W band to measure each galaxy's structural parameters in the rest-frame optical, additionally we use only the blank regions of each field, masking out clustered and strongly lensed regions in cluster fields. Table 1 of \citet{Conselice_2024} gives the total unmasked area and $5\sigma$ depths of each band in each field used in this study.

\subsection{ERS and GO data}
This work makes use of public \textit{Early Release Survey} (ERS) and \textit{General Observer} (GO) data taken from \textit{Cosmic Evolution Early Release Science Survey} (CEERS, PID: 1345, PI: S. Finkelstein, see also \citet{Bagley_2023_CEERS}), the \textit{Great Observatories Origins Deep Survey South} (GOODS-South), a field part of the \textit{JWST Advanced Deep Extragalactic Survey} (JADES, PID:1180, PI: D. Eisenstein, \citet{Eisenstein_2023}) which was released publicly as JADES DR1 \citep{Rieke_2023}, the \textit{Grism Lens Amplified Survey from Space survey} (GLASS, PID: 1324, PI: T. Treu, \citet{Treu_2022}), the \textit{Next Generation Deep Extragalactic Exploratory Public Survey} (NGDEEP, PID: 2079, PIs: S. Finkelstein, Papovich and Pirzkal, \citet{bagley_2023_NGDEEP}) and SMACS-0723 (PID: 2736, PI: K. Pontoppidan, \citet{Pontoppidan_2022}).

We incorporate two HST ACS/WFC bands into our CEERS data (F606W and F814W) to account for the vacancy of the F090W filter from CEERS observations. These bands are taken from observations over the \textit{Extended Groth Strip} (EGS: \citet{Groth_1994}), as part of the \textit{Cosmic Assembly Near-infrared Deep Extragalactic Legacy Survey} (CANDELS, \citet{Grogin_2011}; \citet{Koekemoer_2011}). Similarly we also incorporate F606W and F814W data into our JADES and NGDEEP data, taken as part of the most recent mosaic (v2.5) from the Hubble Legacy Fields team \citep{Illingworth_2016, Whitaker_2019}.

\subsection{PEARLS data}
This work also makes use of NIRCam imaging taken as part of the PEARLS GTO survey. In all, we use data taken from four PEARLS fields; one blank field and three targeted gravitationally lensed cluster field. The targeted blank field is the \textit{North Ecliptic Pole Time Domain Field} (NEP-TDF, \citet{Jansen_&_Windhorst_2018}), consisting of four 'spokes' orientated at 90 degree intervals, each consisting of a pair of overlapping NIRCam pointings, giving us 8 pointings in total.

The cluster fields we utilise consist of Clio (GAMA 100033 in the GAMA Galaxy Group Catalog v6+, \citet{Robotham_2011}), El Gordo (ACT-CL J0102-4915 in the Atacama Cosmology survey \citet{Marriage_2011, Menanteau_2012}) and MACS J0416.1-2403 (Hereafter called MACS-0416). Each field has data in 8 NIRCam bands (F090W, F115W, F150W, F200W, F277W, F356M, F410M and F444W), apart from Clio which lacks F115W and F410M data.   For full details of the PEARLS program please see \citet{windhorst_2023} and \citet{Frye_2023}.

\subsection{JWST NIRCam Data Reduction}
We use a modified version of the official JWST reduction pipeline to uniformly reduce the JWST imaging. This follows a similar process to that used in \citet{Ferreira_2022_disk_panic, Adams_2023, Austin_2023} and is identical to that used in other papers in the EPOCHS series. A more detailed account of this pipeline can be found in \citet{Adams_2024}, which is summarised below.

In our data reduction, we use version 1.8.2 of the official STScI JWST pipeline\footnote{https://github.com/spacetelescope/jwst} \citep{Bushouse_2022} and Calibration Reference Data System (CRDS) v1084, which contains the most up-to-date NIRCam calibrations with updated in flight long wavelength flat fielding which drastically improves depths across a single pointing, up to $\sim 0.2$ dex on average in the F444W filter, compared to CRDS v0995. To account for `wisps' (short wavelength artefacts caused by reflections off the telescopes top secondary mirror support strut) we subtract templates of these wisps from the F150W and F200W imaging, between stage 1 and 2 of the pipeline. To account for 'claws' in our imaging (artefacts from nearby bright foreground stars), we mask out the regions affected post reduction.

After stage 2, we apply Chris Willott's $1/f$ noise correction\footnote{https://github.com/chriswillott/jwst}. To allow for quicker background subtraction assessment and fine-tuning, we bypass the sky subtraction step in stage 3 by performing background subtraction on individual 'cal.fits' frames. This consists of an initial flat background subtraction followed by a further 2D background subtraction using \texttt{photutils} \citep{Bradley_2024}. After stage 3, we align our F444W imaging with the Gaia-DR3 (Data Release 3, \citet{Gaia_2018, Gaia_2023}) derived World Co-ordinate System (WCS) before matching all remaining filters to this WCS, all using the \texttt{tweakreg} part of the DrizzlePac\footnote{https://github.com/spacetelescope/drizzlepac} python package. Finally, we pixel-match to the F444W image using the astropy \texttt{reproject}\footnote{https://reproject.readthedocs.io/en/stable/} \citep{Hoffman_2021} to make drizzled images with a pixel scale of $0.03"$/pixel.

\section{Catalogue and Sample selection}
\label{sec:Catalogue selection}
In this section, we briefly summarise the procedure we use to create our catalogue and select our high-z candidates that are included in our final sample. Full details can be found in \citet{Conselice_2024}.

\subsection{Catalogue creation}

We use the code \texttt{SExtractor} \citep{Bertin_&_Arnouts_1996} to identify and take photometry measurements of sources in each field and pointing. We make use of forced aperture photometry in all photometric bands on sources detected in the inverse-variance weighted stack of the NIRCam F277W, F356W and F444W images to reliably identify faint sources, and create an initial photometric catalogue. This includes band fluxes and magnitudes in both apertures with a radius equal to the Kron radius (``FLUX\_AUTO'' and ``MAG\_AUTO'' e.g.) as well as using a 0.32 arcsecond diameter aperture (``FLUX\_APER'' and ``MAG\_APER'' e.g.). To correct for any flux that extends beyond the aperture being used to measure, we use the point spread function (PSF) using the simulated WebbPSF PSF's \citep{Perrin_2014}. Our aperture diameter, of 0.32 arcseconds, was chosen to enclose the brightest 70-80\% of the flux of a point source, avoiding a large amount of contamination from neighbouring sources, minimizing the aperture correction, and reducing the dependence of using PSF models that are potentially uncertain.

Since \texttt{SExtractor} requires all images to be on the same pixel grid, we use \texttt{photutils}  \citep{Bradley_2024} to perform forced photometry using the same 0.32 arcsecond aperture on HST imaging, which is on a different pixel gird to JWST imaging.

We create masks for our images by eye, covering diffraction spikes within the image, image edges ($\sim 50-100$ pixel border around the detector edges), large foreground galaxies, the cross pattern between the SW detectors and other artifacts (such as `wisps', `claws' and `snowballs'). We calculate depths, by placing random apertures with diameters of 0.32 arcseconds into empty regions as defined by the segmentation map created by \texttt{SExtractor} and our own image masks, in each band. These apertures are placed $\geq 1''$ from the nearest source, to ensure there is no overlap.

We use the flux measurements of the nearest 200 apertures of each source, to calculate the local depth for the source. We do this for each source in every filter. Using these local depths, we calculate the Normalised Median Absolute Deviation (NMAD) of the fluxes measured in each aperture, which corresponds to the $1\sigma$ flux uncertainty, which we convert into a $5\sigma$ depth. The total unmasked area used for each field in this study, along with their $5\sigma$ depths, can be found in Table 1 of \citet{Conselice_2024}.

\subsection{High-z selection}

For our analysis in this work, we select a sample of high-redshift objects using a selection criteria based primarily on photometric Spectral Energy Distribution (SED) fitting performed using the fitting code \texttt{EAZY-py} \citep{Brammer_2008}. Our aim is to select a robust sample of galaxies above z $\geq 6.5$, where the Lyman break is within the NIRCam F090W filter.

We make use of the standard ``tweak\_fsps" templates, produced using the flexible stellar population synthesis (FSPS) package \citep{Conroy_&_Gunn_2010}, along with set 1 and 4 from the bluer \citet{Larson_2023} template set (``fsps\_larson"). These additional templates have bluer rest-frame UV colors than the default templates, as well as stronger emission lines, both of which have been observed in high-redshift galaxies \citep[e.g.][]{Finkelstein_2022, Withers_2023, Austin2024, Nanayakkara_2024, Cullen_2024}. These templates have been shown to improve the accuracy of photo-\textit{z} estimates at high redshifts by \citet{Larson_2023_b}.

We initially run \texttt{EAZY-py} twice, one where the redshift prior is free ($0 \leq z \leq 25$), as well as a low redshift run with an upper redshift limit of $z \leq 6$, allowing us to compare the goodness of fit for both runs. This is required for our selection criteria, and has previously been performed for high-z galaxy identification in the past \citep{Hainline_2024}.

Our selection criteria for high redshift galaxies are as follows.
\begin{enumerate}
    \item We require there to be at least 1 photometric band bluewards of the Ly$\alpha$ break at $\lambda_{\mathrm{rest}} = 1216$\AA, giving us a lower limit of $z \simeq 6.5$ in most of our fields.
    \item Any filter bluewards of the Ly$\alpha$ break must have a detection of $\leq 3\sigma$ in those bands.
    \item The 2 bands directly redward of the Ly$\alpha$ break must have a detection of $5\sigma$, and those further redward must have a detection of $\geq 2\sigma$, with exception of the NIRCam medium-bands (e.g. F335M).
    \item The integral of the \texttt{EAZY-py} redshift probability density function (PDF) must be greater than 60\% within 10\% of the maximum likelihood to ensure the majority of the posterior is located within the primary peak.
    \item We require the best-fitting \texttt{EAZY-py} SED fitting to satisfy $\chi^{2}_{\rm red} < 3(6)$ for a robust (good) fit.
    \item To ensure a higher-redshift solution is statistically more probable, we require a difference of $\Delta\chi^{2}_{\rm red} \geq 4$ between the low-z and the high-z solutions.
    \item To remove Milky Way brown dwarf contaminants, we require the best-fitting high-z solution and the best-fitting brown dwarf template to have a $\Delta\chi^{2}_{\rm red} \geq 4$. This is only applied if the FLUX\_RADIUS parameter (from \texttt{SExtractor}) in the F444W band is smaller than the FWHM of the PSF. To remove these contaminants, we fit a synthetic \texttt{Sonora Bobcat} template \citep{marley_2021} using a simple $\chi^2$ minimisation. For more information we direct the reader to \citet{Harvey2024}.
    \item To avoid the selection of over-sampled hot pixels from the LW detectors as F200W dropouts, we require the FLUX\_RADIUS parameter to be $\geq1.5$ pixels in the long-wavelength wideband NIRCam bands (F277W, F356W and F444W).
\end{enumerate}

\begin{figure*}
    \centering
    \includegraphics[width=1\linewidth]{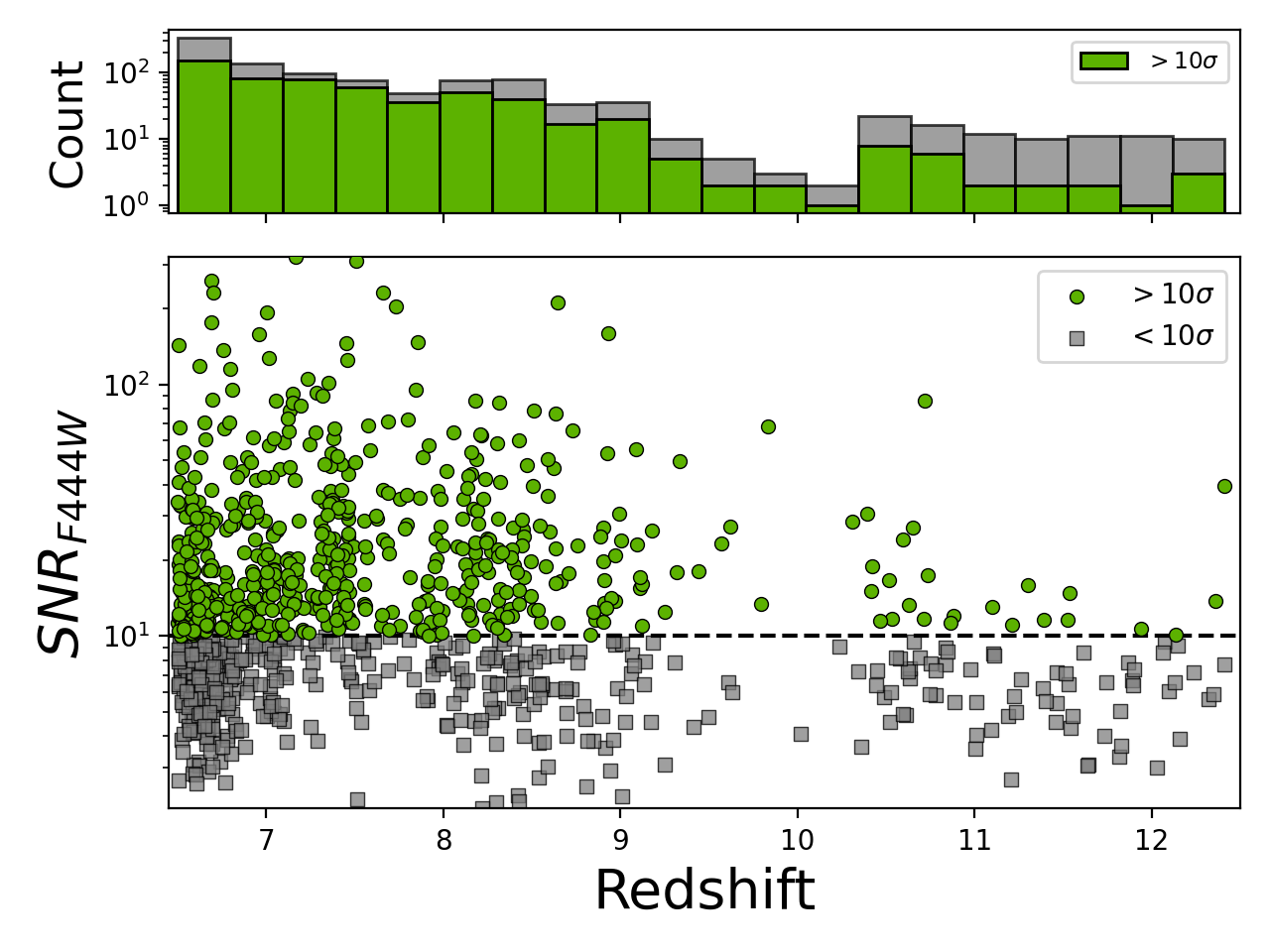}
    \caption{Redshift distribution of the EPOCHS v1 sample, with our $10\sigma$ SNR cut used in this work, in the form of the black dashed line. Gray squares are galaxies not included in this work due to having a SNR below $10\sigma$, while green circles are those that are included in this work. The histogram shows the redshift distribution both before and after the SNR cut. The redshift axis is shared between both plots.}
    \label{fig:z_vs_SNR}
\end{figure*}

To account for surface brightness profile modelling often incorrectly identifying the boundaries of a faint galaxy with a low signal-to-noise ratio (SNR), we make a cut of SNR $> 10$ of aperture flux in the F444W filter to eliminate less reliable models entering our final sample, giving us an initial sample of 567 galaxies. The redshift distribution of our sample, along with the SNR cut are shown in Fig. \ref{fig:z_vs_SNR}. We note that there is a dip in the number of objects found  at $z \sim 10$, this is likely due to photo-z scattering as a result of the placement of the observed wavelength of the Lyman break. At this redshift, the break falls in a gap between the F115W and F150W NIRCam filters, leading to an object's photometric redshift being slightly overestimated or underestimated. As a result of this, when we define our redshift bins, we combine bins containing objects at $z \sim 10$ with objects at $9.5 < z < 11.5$.

\begin{figure*}[p]
    \centering
    \includegraphics[width=0.9\linewidth]{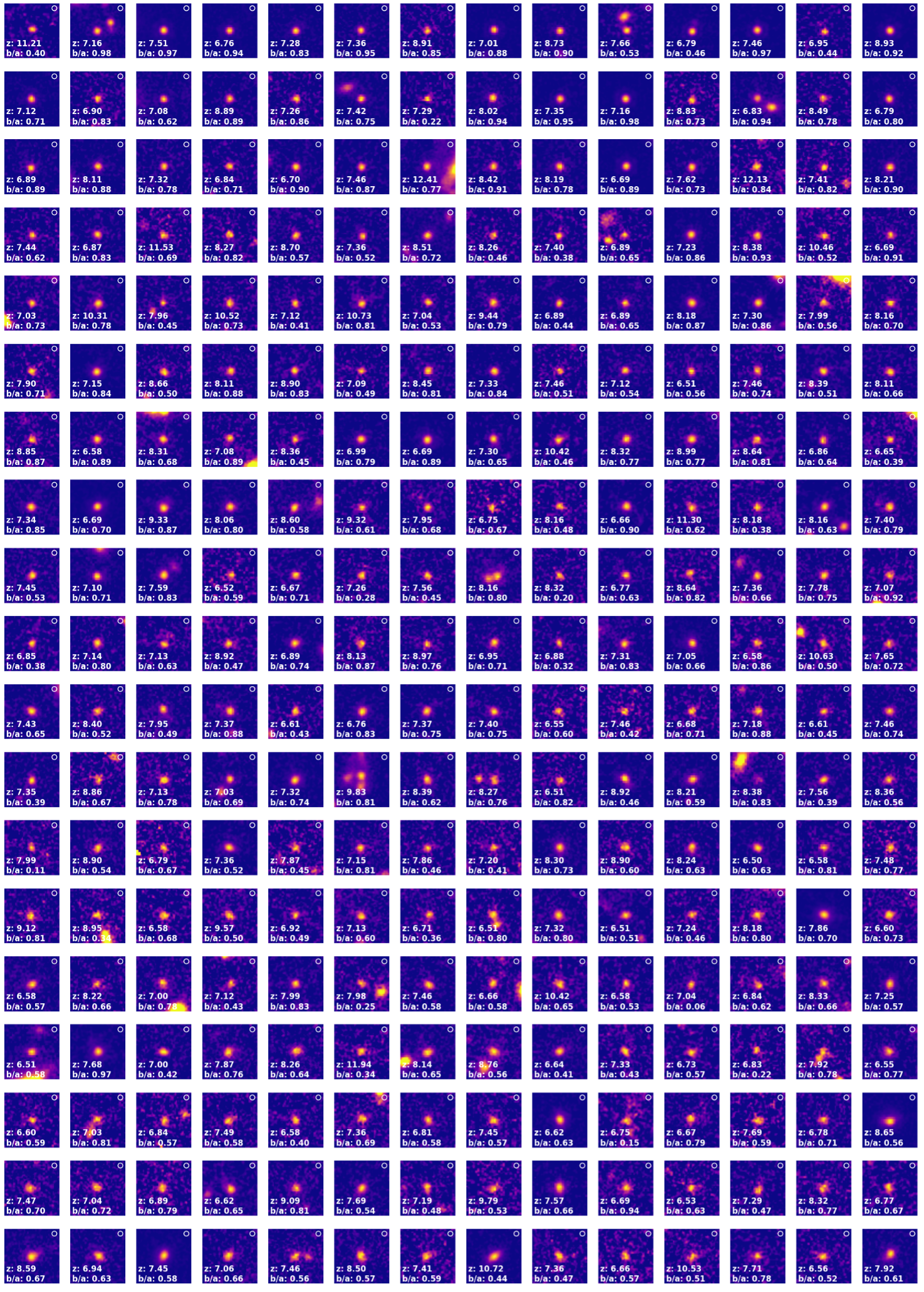}
    \caption{\textbf{(a)} Mosaic of $1.5 \times 1.5$ arcsec F444W cutouts of all objects in our sample that pass our selection criteria as outlined in section \ref{sec:GALFIT criteria}, sorted by their \texttt{GALFIT} measured sizes, smallest to largest. Each cutout displays the FWHM of the F444W PSF in the top right corner. Each objects redshift and axis-ratio are displayed in the bottom left.}
    \label{fig:good_cutouts}
\end{figure*}

\begin{figure*}[p] 
    \ContinuedFloat 
    \centering
    \includegraphics[width=0.9\textwidth]{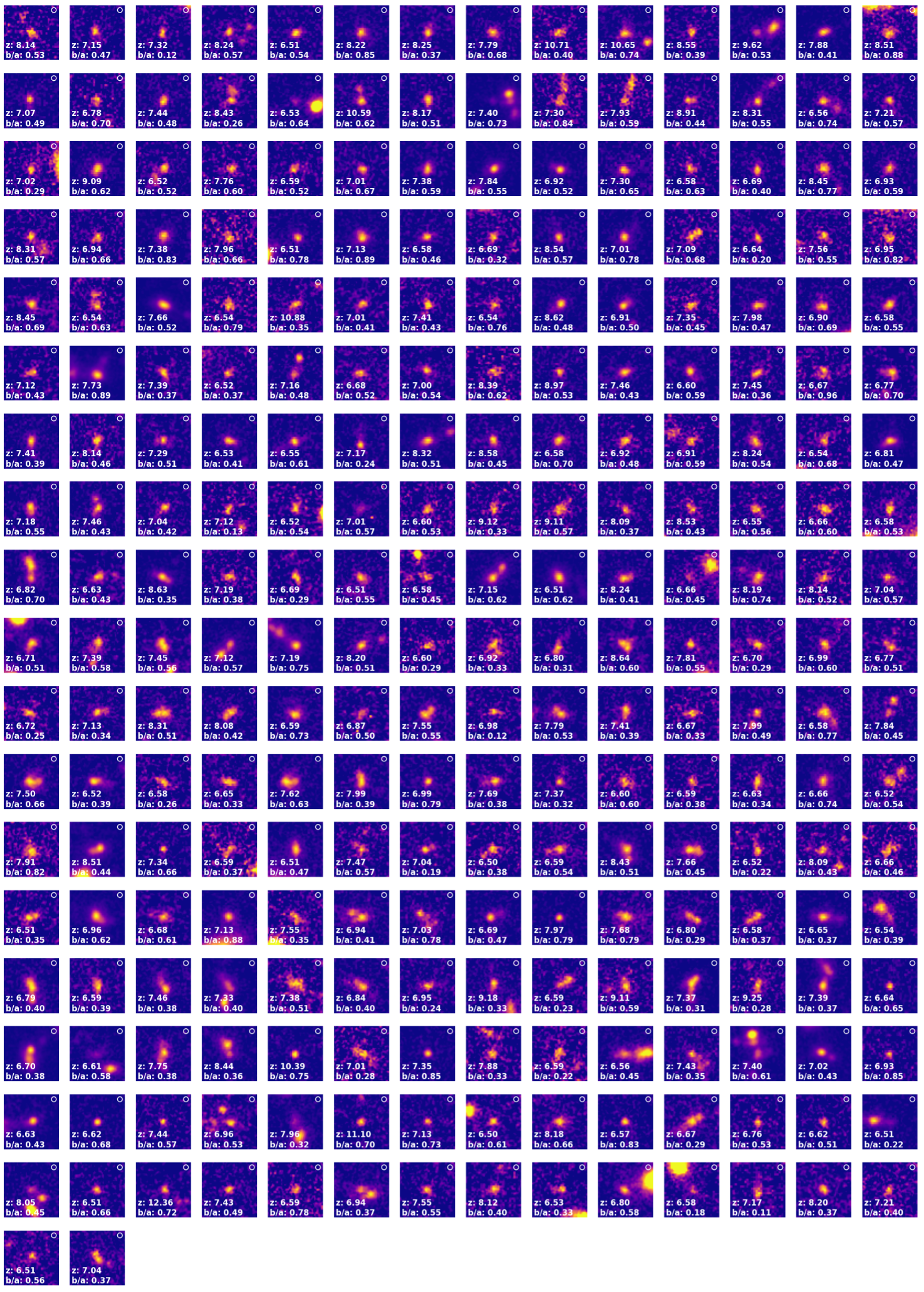} 
    \caption{\textbf{(b)} Mosaic of $1.5 \times 1.5$ arcsec F444W cutouts of all objects in our sample that pass our selection criteria as outlined in section \ref{sec:GALFIT criteria}, sorted by their \texttt{GALFIT} measured sizes, smallest to largest. Each cutout displays the size of the F444W PSF in the top right corner. Each objects redshift and axis-ratio are displayed in the bottom left.}
\end{figure*}


\begin{figure*}[p]
    \centering
    \includegraphics[width=0.9\linewidth]{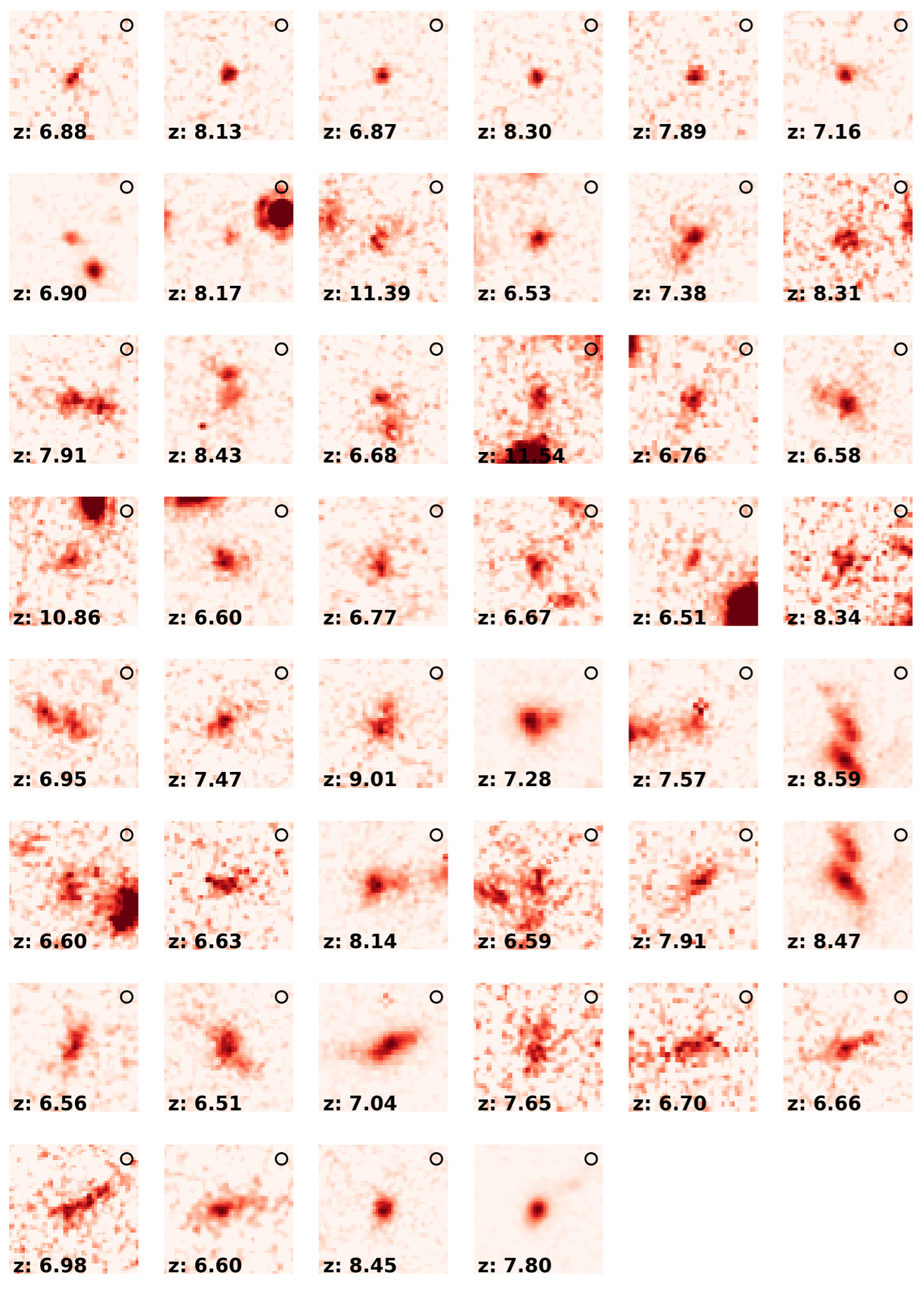}
    \caption{Mosaic of $1.5 \times 1.5$ arcsec F444W cutouts of all objects in our sample that failed our selection criteria as outlined in section \ref{sec:GALFIT criteria}. Each cutout displays the size of the F444W PSF in the top right corner.}
    \label{fig:bad_cutouts}
\end{figure*}

\subsection{Stellar Mass Measures}

We also measure based on the SEDs of our sample and their photometric redshfits, the stellar masses of our sample using the code \texttt{Bagpipes} \citep[][]{Carnall2018}.  A large amount of detail for how these masses are calculated are include in \cite{Harvey2024} and \cite{Conselice_2024}, although we give a summary of our methods here for completeness.

We use the \texttt{Bagpipes} code with a range of models, priors and star formation histories, which can have a large impact on derived galaxy properties \citep[e.g.][]{Carnall2018}. Through our previous analysis \citep[][]{Harvey2024} we have many different possible stellar masses to use under different assumptions in parametric and non-parametric methods.  

However, in this paper we present results of our fiducial \texttt{Bagpipes} run. These stellar masses are calculated using a log-normal star formation history with logarithmic priors on age, dust extinction, and metallicity. This log-normal star formation history was chosen to represent the predicted 'rising' star formation rate of high$-z$ galaxies \citep[e.g.][]{Adams_2024}.  As we and many others have found that the quantification of dust, metallicity, and age are difficult to constrain accurately based on photometry alone, our prior leads to at times low ages, low dust extinction and low metallicity, which is predicted by simulations and confirmed by spectroscopy. We assume Calzetti dust emission,  ISM extinction and stellar population models from \cite{Bruzual2003}. To calculate these masses we use an informative redshift prior based on our EAZY results, with a Gaussian centered on the median of the EAZY redshift posterior, and standard deviation based on the average of the 16 and 84th percentiles of the PDF, and capped at $\pm3\sigma$. This fiducial model was compared to the masses measured with a "delayed" star formation history and a "continuity bursty" star formation history. In both cases, through a goodness of fit $\chi^2$ test, the fiducial log-normal star formation history was preferred in the majority of cases.

\section{Morphological fitting}
\label{sec:morphological fitting}
In this work, we use \texttt{GALFIT} to model 2D light profiles \citep[][]{2002AJ....124..266P, Peng2010}, and \texttt{Morfometryka} to measure non-parametric parameters \citep{ferrari2015morfometrykanewway}

\subsection{GALFIT}
\label{sec:GALFIT}
We utilise \texttt{GALFIT} \citep[][]{2002AJ....124..266P, Peng2010} to fit a single Sérsic light profile to each galaxy. \texttt{GALFIT} is a least-squares-fitting algorithm, which utilises a Levenberg-Marquardt algorithm to find an optimum surface brightness profile for each galaxy, determined via a goodness-of-fit test to find the best fitting model through the reduced chi squared minimisation, $\chi^{2}_{\rm min}$. The $\chi^{2}_{\rm min}$ is given by:

\begin{equation}
    \chi^{2}_{\rm min} = \frac{1}{N_{\rm DOF}} \sum^{nx}_{x = 1}{\sum^{ny}_{y = 1}{\frac{ \left( f_{\rm data} \left( x, y \right) - f_{\rm model} \left( x, y \right) \right)^{2}}{\sigma \left( x, y \right)^{2}}}} 
    \label{equ:reduced chi squared}
\end{equation}
summed over all $nx$ and $ny$ pixels where $N_{DOF}$ is the number of degrees of freedom. As can be seen from equation \ref{equ:reduced chi squared}, \texttt{GALFIT} requires a data image, from which a given galaxy's surface brightness is measured, $f_{\rm data}$ and a sigma image, or error map, $\sigma$ which gives the relative error at each pixel within the image, which is then used to calculate a model image $f_{\rm model}$.

In this work, we run \texttt{GALFIT} for NIRCam's F444W band, to best probe the rest frame optical to minimise the effect of morphological K-correction, as the morphological structure for an observed galaxy changes as a function of wavelength, both qualitatively and quantitatively \citep{Taylor-Mager_2007}.

From \texttt{GALFIT}, the calculated Sérsic profile follows the form:
\begin{equation}
    I \left( R \right) = I_{e} \exp \left\{ -b_{n} \left[ \left( \frac{R}{R_{e}} \right)^{1/n} - 1 \right] \right\}
\end{equation}
where $n$ is the Sérsic index, which controls the shape of the light profile, $I(R)$ is the intensity of light at a distance $R$ from the galaxy's centre, $R_{e}$ is the half-light radius, where 50\% of light is contained within and $I_{e}$ is the intensity of light at the half-light radius \citep{1963BAAA....6...41S, Cioti_1991, Caon_1993}. $b_{n}$ can be approximated to be $b(n) \approx 2n - \frac{1}{3} + \frac{4}{405n} + \frac{46}{25515{n^2}}$ \citep{ciotti_1999}. \texttt{GALFIT} gives us the best fit values for each term, along with the error bars, which are calculated through the same method, a full description of the \texttt{GALFIT} error calculation can be found in \citet{2002AJ....124..266P}.

\begin{figure*}
    \centering
    \includegraphics[width = \linewidth]{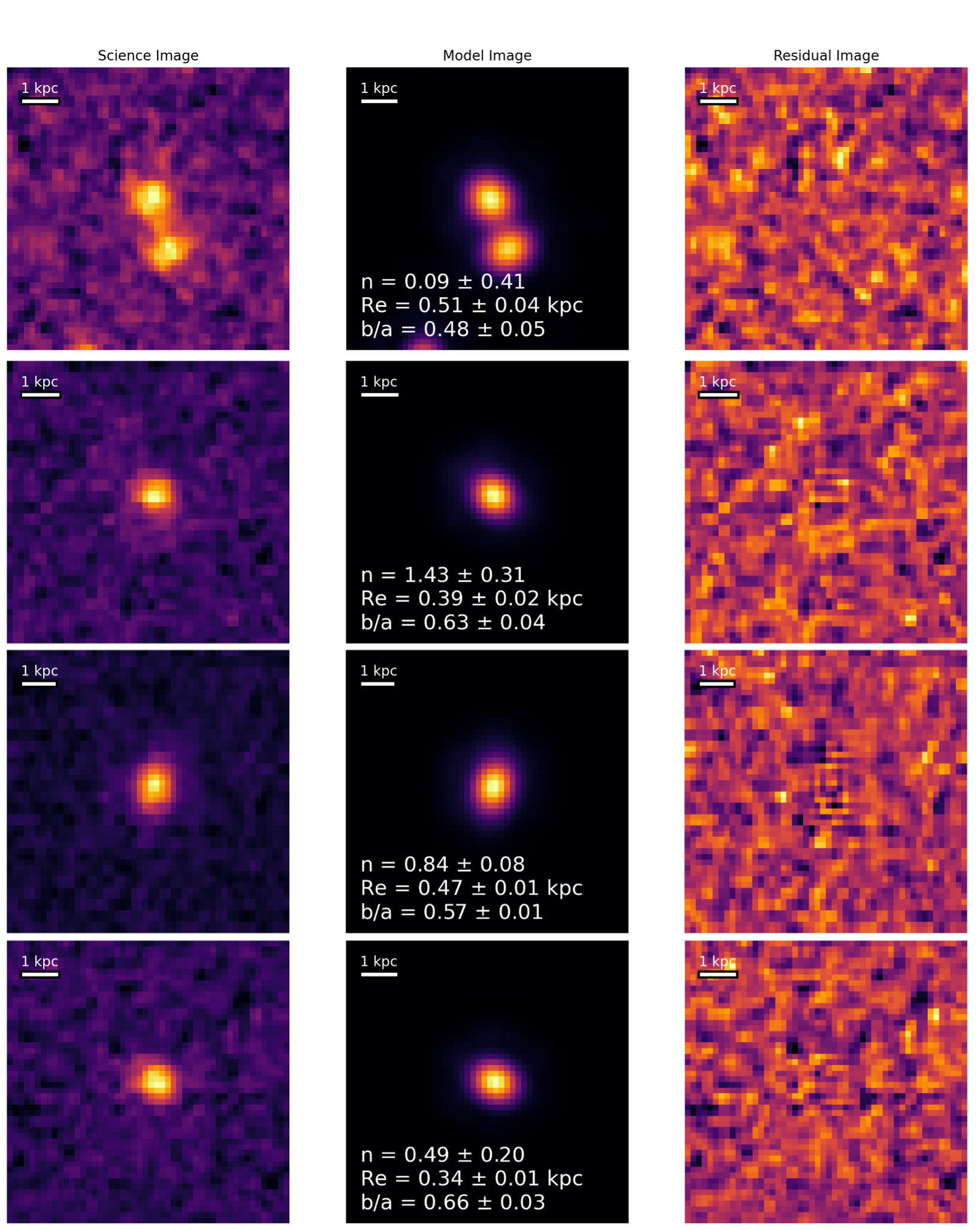}
    \caption{Example runs of \texttt{GALFIT} from galaxies used in this work. Each example displays the original cutout image (left panel), single S\'ersic model (middle panel) and residual (right panel), with the exception of the top panel which shows an example where neighbouring galaxies within the cutout have also been modelled with their own single S\'ersic profile. Measured values are taken from the S\`ersic profile of the central galaxy in each cutout. White bars correlate to 1 pkpc at each objects photometric redshift.}
    \label{fig:Galfit_outputs}
\end{figure*}

We run a custom \texttt{GALFIT} pipeline in order to model single-component S\'ersic fits and measure their parameters. This is done through the following.

\begin{enumerate}
    \item \textbf{Cutouts:} We create cutouts for each object of $100 \times 100$ pixels ($3 \times 3$ arcsec). This creates a cutout which encompasses the entire source and ensures the surface brightness profile of the galaxy is enclosed within. 
    Following the same dimensions, we also make cutouts for the sigma image that \texttt{GALFIT} is required to run. For this image we take a cutout of the ‘ERR’ extension of our images, which is a measure of the RMS noise of the image. The addition of this sigma image allows \texttt{GALFIT} to give a relative weight to the pixels during the fit, if an image is not provided, \texttt{GALFIT} will attempt to make one itself, however the data must be in a specific form for this to be done properly. To avoid this, and for the best results, we provide our own.
    \texttt{GALFIT} further requires the inclusion of a mask, which prevents other objects in the image from being included in the S\'ersic model being made. To make these masks, we use the SExtractor segmentation maps, and exclude the object we want modelled, leaving us with a segmentation map of objects to be excluded from the fitting process.
    \item \textbf{Input parameters:} \texttt{GALFIT} requires initial estimates for each parameter, which it then iterates upon to find the best fitting model. The parameters required are the position of the $x$ and $y$ coordinates of the object being modelled in the image, total magnitude, half-light radius, axis ratio, position angle and S\'ersic index. Similarly to \citet{Kartaltepe_2023, ormerod_2024}, we use the \texttt{SExtractor} catalogue for our initial parameter estimates. These catalogues do not contain values for the S\'ersic index, so we estimate this to be n = 1 initially and allow the fitting process to determine the best fitting value. We find that altering the initial value of the S\'ersic index does not significantly change the output, since these input parameters are just initial estimates and do not need to be particularly accurate. We further provide constraints on the model to ensure that the correct object is being modelled, this is done by constraining the image position to be within 5 pixels of the initial estimate.
    \item \textbf{PSF:} \texttt{GALFIT} requires the appropriate PSF. We obtain our PSF models using the package \texttt{WebbPSF} \citep{Perrin_2014}, and resample to our pixel scale. When dealing with objects which are comparable in size to the PSF, it is important we ensure our PSF is reliable, for this reason we test the performance of different PSF models in Appendix \ref{sec:PSF_comparison}.
\end{enumerate}

\subsection{\rm \texttt{Morfometryka}}
 \label{sec:mfmtk}

We used the algorithm \texttt{Morfometryka} \citep{ferrari2015morfometrykanewway} for non parametric measurements of the photometry and morphological parameters. In the current implementation, given the galaxy stamp and PSF, the code interactively estimates the background by sigma-clipping sources,  segments the sources and deblends them if needed. In the segmented region it measures the standard geometric properties (center, axis length and ratio, position angle) using image moments; these values are used to define equally spaced similar ellipses  based on which basic photometry is performed. The Petrosian radius, $R_p$, is defined where the ratio $\langle I(R) \rangle /I(R)=5$ (the ratio of the mean intensity inside the ellipse and the intensity over the ellipse) is reached \citep{Petrosian1976ApJ...209L...1P}.  The photometric measurements are truncated at $2R_p$ and an elliptical  region with the same shape as the galaxy and with major axis of 2$R_p$ is defined; later measurements are made in this Petrosian region. Other sources from segmentation and deblending are kept masked out in this region. Similarly, the Kron radius \citep{Kron1980ApJS...43..305K}, i.e. the intensity weighted average radius, is measured by two different methods. Both $R_p$ and $R_k$ are efforts to ascertain some estimate of the galaxy size in a robust way, insensitive to noise and given that we do not observe the galaxy outer limits.

Standard morphometric parameters  are calculated as in \citet{Conselice_2003, ferrari2015morfometrykanewway} and references therein.  From the luminosity growth curve, the radii $R_f$ containing a fraction $f$\% of total light, $f\in \{10,20,\ldots100\}$, are measured. From them the concentrations
\begin{eqnarray}
C_1=5\log\frac{R_{80}}{R_{20}} \qquad {\rm and} \quad  C_2=5 \log\frac{R_{90}}{R_{50}}
\end{eqnarray}
are calculated. 
The standard asymmetry $A$ -- the normalized sum of the residual between the original image and the 180$^\circ{}$ rotated version -- is measured, choosing the center that minimizes its value.

\section{Sample selection}
\label{sec:sample selection}
We choose galaxies for our analysis by making selections based on the \texttt{GALFIT} outputs for each object. Doing so checks the goodness of fit, and removes objects that have unphysical output parameters.
While \texttt{GALFIT} does output a $\chi^{2}$ value, we do not use this when making our selection, as \texttt{GALFIT} only uses this value to determine when it has reached the best fitting, and a high $\chi^{2}$ value on a single S\'ersic model could also be due to a more peculiar structure.
Instead, we make our selections based on the output parameters, as well as the Residual flux fraction (RFF) values.
\subsection{GALFIT criteria}
\label{sec:GALFIT criteria}
In order for a galaxy to pass as a good fit, we require the \texttt{GALFIT} single S\'ersic model to pass the following criteria:

\begin{enumerate}
    \item $0.01 < R_{e} $ (pixels) $ < 50$. This requires that the half light radius can not take up more than half the cutout, and also can not be so small that it is unphysical.
    \item $0.05 < n < 10$. This removes unphysical outputs of s\'ersic indices.
    \item $(b/a) > 0.01$. This requires the best fit axis ratio to not be unphysical. This is particularly important for fainter objects, whose fits sometimes converge on a very small axis ratio \citep{vanderWel_2012}.
\end{enumerate}

We discuss how many objects pass these criteria in section \ref{sec:final_sample} and what aspect galaxies we remove are failing within these criteria. We also discuss these `rejected' galaxies in more detail later. 

\subsection{Residual Flux Fraction}
We make further cuts to our sample after calculating the RFF on those objects which passed our checks as detailed in Section \ref{sec:GALFIT criteria}. The RFF gives us a measure of how well the model created by \texttt{GALFIT} matches our original image, by giving us a measure of the signal in the residual image, that can not be explained by background fluctuations. Similar to \citet{Margalef_2016}, we define the RFF to be:

\begin{equation}
    RFF = \frac{\sum{\left| I_{j,k} - I^{\rm GALFIT}_{j,k} \right|} - 0.8 \sum{\sigma_{B j,k}}}{\rm FLUX\_AUTO}
    \label{eq:RFF}
\end{equation}

\noindent where $I$ is the NIRCam image that is input to \texttt{GALFIT}, $I^{GALFIT}$ is the model image created by \texttt{GALFIT}, $\sigma_{B}$ is the background of the RMS image, and FLUX\_AUTO is the flux of the galaxy measured through SExtractor. The 0.8 factor ensures that the expected value for the RFF is 0 for a Gaussian noise error image \citep{Hoyos_2011}. The RFF is calculated within the radius of 2 FLUX\_RADIUS, which is a half-light radius measurement measured with \texttt{SExtractor}.
We calculate the background term $\sigma_{B j,k}$ by following the same method used in \citet{Margalef_2016}, where we assume that:
\begin{equation}
    \sum_{(j,k) \in A}\sigma_{B j,k} = \bold{N} \left< \sigma_{B} \right>,
    \label{eq:background}
\end{equation}
where $\left< \sigma_{B} \right>$ is the mean value of the background for the whole image. We calculate this by placing apertures on blank regions of the sigma image we use as an input to \texttt{GALFIT} (see section \ref{sec:GALFIT}) and finding the mean of these aperture regions. The value $N$ is the number of pixels within the radius we are calculating the RFF within.
In order to utilise the RFF to further make cuts to our sample, we decide to have a cutoff RFF value of $0.5$, keeping anything that has RFF $< 0.5$. This allows us to remove objects whose light has been either very over-accounted for or very under-accounted for. 

Should any galaxy fail to meet these criteria, we pass the object through \texttt{GALFIT} again, this time holding the S\'ersic index at a fixed value of $n = 0.05$. We discuss the effect of holding the S\'ersic index in Appendix \ref{sec:holding n}. If this model continues to fail to meet the required criteria, we then reject it from our final sample.

Again, we discuss how many objects pass our RFF test in section \ref{sec:final_sample}.

\subsection{Final Sample}
\label{sec:final_sample}
We next check each galaxy for any nearby potential neighbours which could interfere with our model outcome, which we would model simultaneously as a separate single S\'ersic surface profile model. We opt to model these neighbours instead of masking them out, since our masks do not always sufficiently mask out the neighbours outermost light, or the mask may inadvertently cover some of the object we are interested in, leading to unreliable models. We do this for 62 objects in our initial catalogue.

From our initial 567 galaxies, 448 galaxies pass our criteria check, leading us to holding the S\'ersic index values of the remaining 119 galaxies. Of those 119 that failed this initial selection, 29 objects failed due to having a half-light radius which was too large or too small, a further 55 objects were deemed to have an unphysical S\'ersic index and finally, out of the remaining failed objects, a further 35 were removed for failing our RFF criteria. No objects were cut based on their best fit axis ratio. Once we create new models for these 119 galaxies, with a held S\'ersic index, a further 73 objects pass our criteria, with us removing the remaining 46 galaxies from our final sample, 10 removed due to their measured half-light radius and 36 due to the RFF.

\begin{figure*}
    \centering
    \includegraphics[width=1\linewidth]{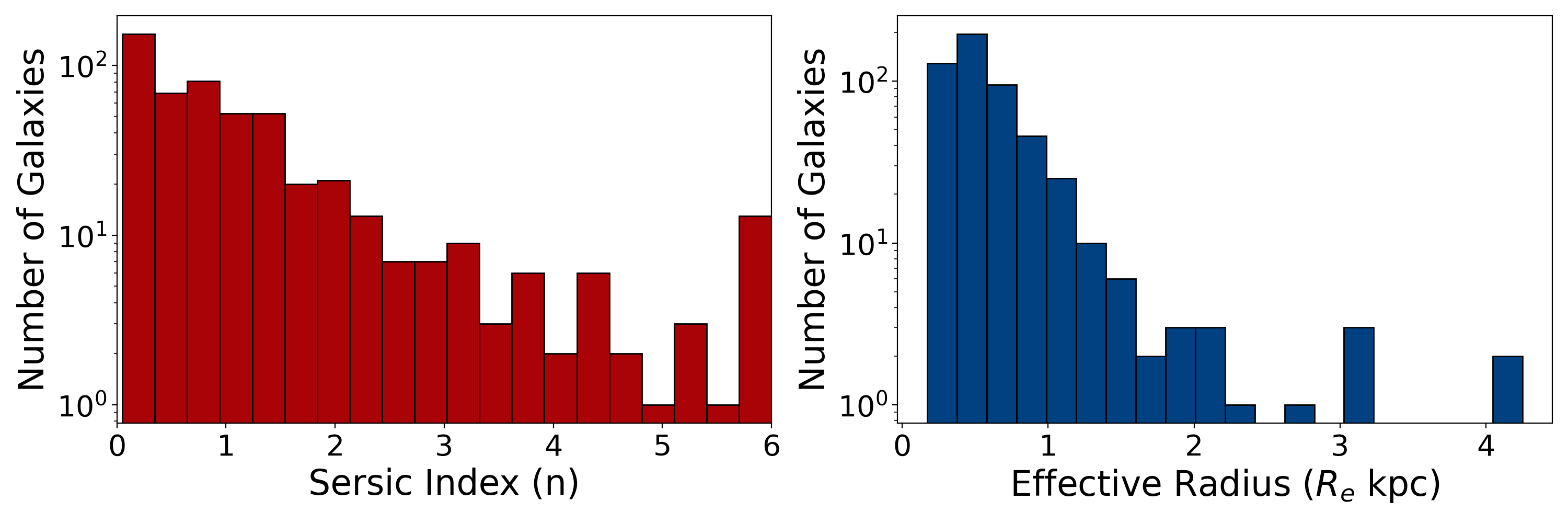}
    \caption{Distributions of different parametric morphological parameters, as measured through \texttt{GALFIT} for our sample. Left panel: distribution of S\'ersic index (n). Right panel: distribution of half-light radii ($R_{e}$) converted into kpc.}
    \label{fig:size_sersic_distribution}
\end{figure*}

At the end of this process we are left with a final sample of 520 galaxies with a single S\'ersic profile model which passes our selection criteria. Within this sample we have 448 galaxies with a S\'ersic index allowed to vary, and 73 with a fixed S\'ersic index of 0.05. Cutouts for each object in our final sample are displayed in Fig. \ref{fig:good_cutouts} (split into panel a and b), whilst those objects that failed our selection criteria can be found in Fig. \ref{fig:bad_cutouts}. Example \texttt{GALFIT} outputs can be seen in Fig. \ref{fig:Galfit_outputs}, and the distribution of our parametric S\'ersic index and half-light radii values are found in Fig. \ref{fig:size_sersic_distribution}.

\section{Results}
\label{sec:results}

\subsection{Size-mass relation}

\begin{figure*}
    \centering
    \includegraphics[width=1\linewidth]{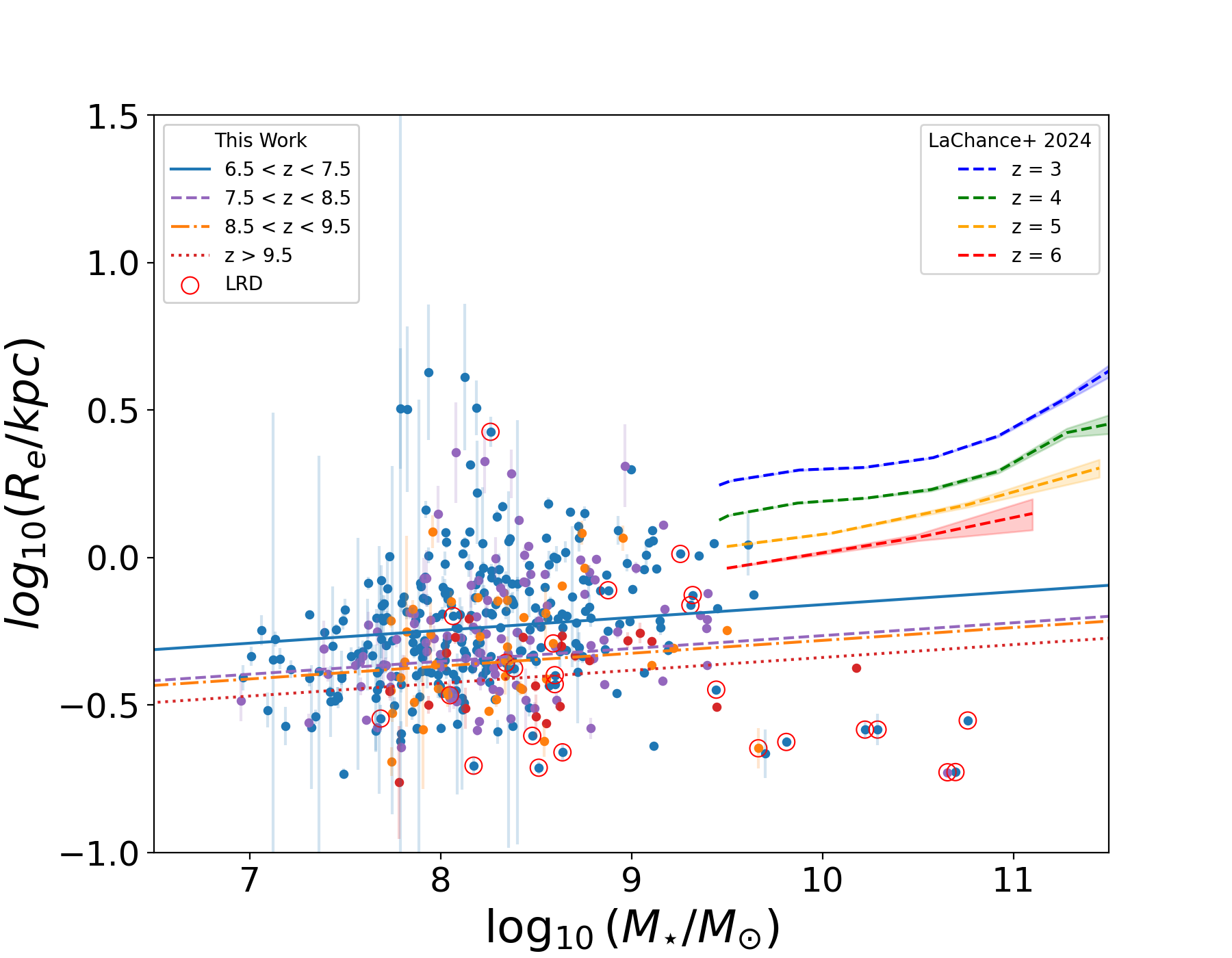}
    \caption{Size-mass relation, with best fit lines for each redshift bin. For each bin, the gradient is fixed based on the best fitting line for the lowest Redshift bin of $6.5 < z < 7.5$. Data points are individual galaxies, with their half-light radii errors as output by \texttt{GALFIT}. Data points are coloured by their corresponding redshift bin, following the same colour scheme as the best fitting lines. Dashed lines in the top right are simulation results taken from the ASTRID simulation \citep{lachance2024evolutiongalaxymorphologyredshift}. Red circled data points are those considered to be a little red dot (LRD), which meet the LRD criteria from \citet{kokorev_2024}.}
    \label{fig:size_mass}
\end{figure*}

As stated in section \ref{sec:morphological fitting}, we measure our galaxy sizes, which we convert into their physical sizes in kpc, through \texttt{GALFIT}. We plot our measured half-light radii and their \texttt{Bagpipes} SED-measured stellar mass for each redshift bin and plot them in Fig. \ref{fig:size_mass}. Our stellar masses are measured with a log-normal star formation history (SFH), as measured through \citet{Harvey2024}. We find a best fit line for our lowest redshift bin ($6.5 < z < 7.5$), as this bin is the most galaxy abundant, and use the gradient found from this fit as a fixed gradient best fit line for the remaining redshift bins. We find that galaxy sizes get progressively smaller and more compact at increasing redshift for a given mass even in the high-redshift regime. Through our size-mass relation, we find galaxies grow in size for a given mass by $\sim 53\%$ in a time interval of just $\sim 400$ Myr. To put this into context, a galaxy of a given mass $\log(M_{*}/M_{\odot}) = 8$ is seen to grow from a size of $\sim 370$ pc to $\sim 570$ pc in this time. The parameters for our best fit lines are found in Table \ref{tab:best_fit_lines}.

We take care to note the inclusion of objects known as `Little Red Dots' (LRDs), a population of red objects with a distinctive `V-shaped' SED. We highlight those objects which meet the LRD criteria from \citet{kokorev_2024} with a red circle in this analysis. We highlight these objects as LRDs have been observed to have characteristic compact sizes and high stellar masses \citep{akins_2024, kokorev_2024, labbe_2023}. 

This trend of smaller, more compact, galaxies of a certain mass with increasing redshift is also seen in simulated data presented by \citet{lachance2024evolutiongalaxymorphologyredshift}. They conduct mock observations through the ASTRID simulation \citep{Bird_2022, Ni_2022} at $3 < z < 6$, based on the CEERS field, which predicts an evolution of the size-mass relation with increasing redshift. While our redshift range is not within the same range in their simulations, we see a similar evolution for a higher redshift range, falling in line with what is seen in simulations presented by  \citet{lachance2024evolutiongalaxymorphologyredshift}.

The galaxies in our sample exhibit stellar masses which are smaller than those used by \citet{ lachance2024evolutiongalaxymorphologyredshift}. In the ASTRID simulation, a mass cut of $\log(M_{*}/M_{\odot}) > 9$ is applied, however it is stated that the simulation finds an overabundance of galaxies with $\log(M_{*}/M_{\odot}) > 9.5$ when compared to what is seen in the CEERS field (by a factor of 1.6-1.9). Conversely, there is an under abundance found for the number of galaxies with masses below the mass cut that are resolved well enough to make morphological fitting reliable. This is argued to be a result of their imaging pipeline, and an indication of some difference in properties between the low mass galaxies in their simulations and those seen in observations. Due to this, our sample contains more lower mass galaxies, at higher redshifts, than what is seen in their simulation.

\begin{table}[h]
    \centering
    \begin{tabular}{c|c|c}
        Redshift & Gradient & $\log_{10}{(R_{e0} [kpc])}$\\
        \hline
        $6.5 < z < 7.5$ & $0.04 \pm 0.02$ & -0.6\\
        $7.5 < z < 8.5$ & - & -0.7\\
        $8.5 < z < 9.5$ & - & -0.72\\
        $z > 9.5$ & - & -0.77\\
    \end{tabular}
    \caption{Best line fit parameters for our size-mass relations in Fig. \ref{fig:size_mass}. All lines have the same gradient, found for our lowest redshift bin $6.5 < z < 7.5$ and held constant for all other bins due to low number statistics.}
    \label{tab:best_fit_lines}
\end{table}

It has been reported that galaxies may experience a break at a mass of $4 \times 10^8 M_{\odot}$ \citep{chamba_2024}. This break has been reported to separate galaxies into two mass regimes, in which lower mass galaxies regulate their sizes more rapidly due to feedback and environmental processes, resulting in a steeper size-mass relation \citep{chamba_2024}. We test this amongst our sample, by splitting into a low mass and high mass regime at $4 \times 10^8 M_{\odot}$. We observe a similarly steeper gradient to our size-mass relation at this mass break, adding support to the notion of less massive galaxies building up their size quicker due to feedback and environmental processes. We discuss this in more detail in Appendix \ref{sec:mass-break}.

\subsection{Size and Sérsic evolution}
\label{sec:re evolution}

\begin{figure}[h]
    \centering
    \includegraphics[width=\linewidth]{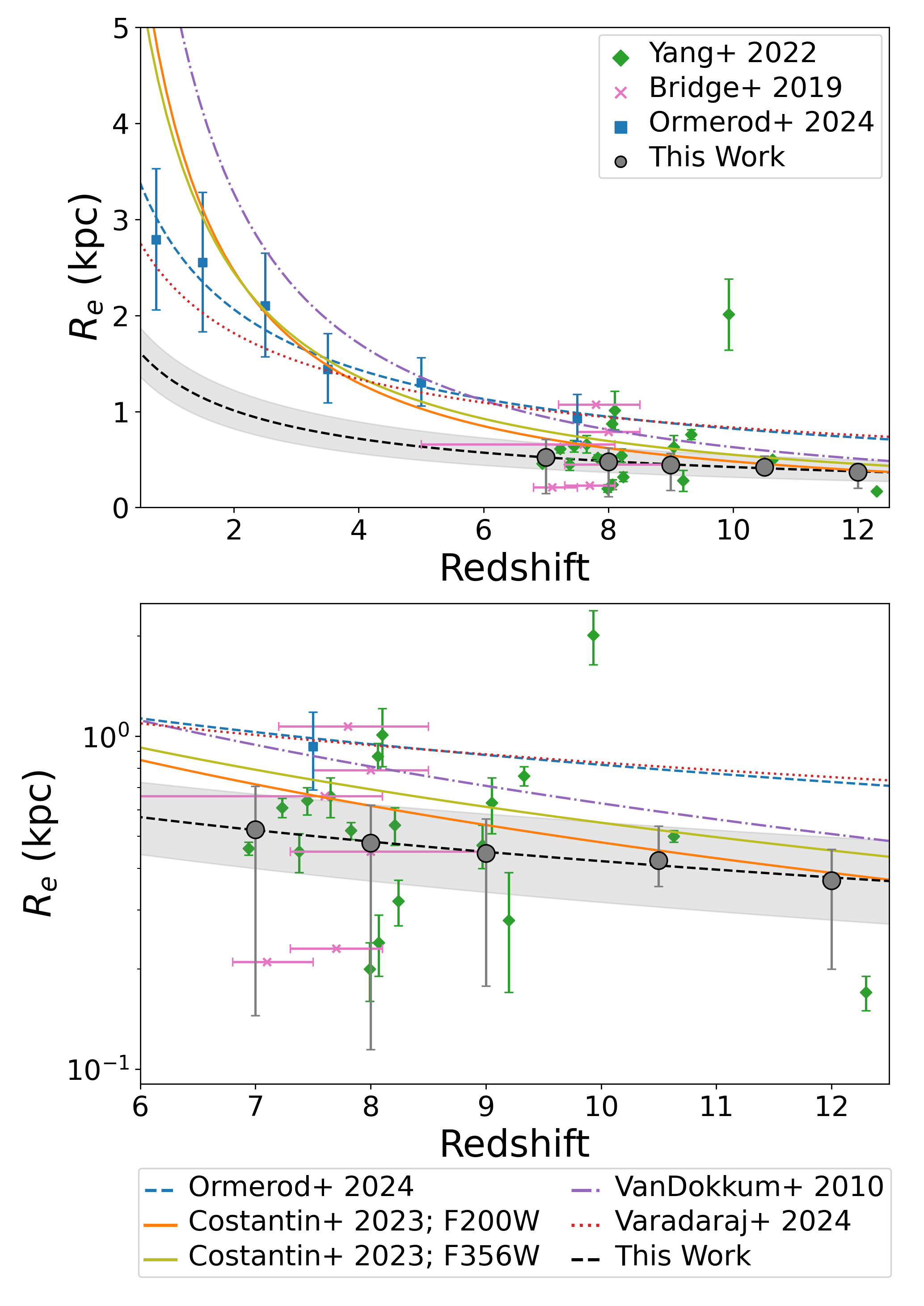}
    \caption{Average galaxy half-light radii for galaxies within our sample. The black-dashed line shows the best fitting power-law, with the form $ R_{e} = \left( 2.12 \pm 0.28 \right) \left( 1 + z \right) ^{-0.67 \pm 0.06}$, with the grey shaded area showing the error of the power-law fit. The grey circular points are the median values of each redshift bin, with the errors being the length of the 84th and 16th percentiles. The previous work in the figure uses JWST data, except for \citet{Bridge_2019} which uses HST data, \citet{VanDockum2010} which uses NOAO/Yale NEWFIRM Medium Band Survey Data and \citet{Costantin2023} which uses simulations to produce mock CEERS observations. For \citet{Bridge_2019} only redshift errors are plotted, as radius errors are not provided.}
    \label{fig:re_evolution}
\end{figure}

Having found evidence for an evolving size-mass relation into the high redshift regime, we next focus on how galaxy sizes and S\'ersic indices evolve into these early epochs. Fig. \ref{fig:re_evolution} plots our average galaxy sizes in different redshift bins, along with a fitted power law to our data with its error and extrapolated power laws from previous works \citep{VanDockum2010, Costantin2023, ormerod_2024, varadaraj2024sizesbrightlymanbreakgalaxies}. Also plotted are the sizes of individual galaxies from previous work, \citet{Yang_2022} from early JWST data, and \citet{Bridge_2019} from HST data. The top panel displays our extrapolated power for lower redshifts that what has been observed, down to a redshift of z = 0, while the lower panel showcases just the observed redshift range with galaxy sizes in log scale.

All other previous work shown utilises JWST data with the exceptions of \citet{VanDockum2010} which uses NOAO/Yale NEWFIRM Medium Band Survey data, \citet{Costantin2023} which utilises TNG50 simulations to produce mock CEERS observations and \citet{Bridge_2019} which uses HST data. 

Our extrapolated best fit power law fit, which can be found in equation \ref{eq:power_law}, follows a similar trend to what is observed at lower redshifts. As redshift decreases we see a predicted increase in galaxy size, with a steeper increase emerging after $z \sim 6$, matching what is observed in size studies at z $< 6$.

The study by \citet{ormerod_2024} similarly analysed the evolution of galaxy sizes, in the rest-frame optical, with increasing redshift, for a lower redshift range of $0.5 < z < 8$ using 1395 galaxies from JWST CEERS data. We plot both their data points and fitted power law to better compare the size evolution of galaxies through JWST data. \citet{ormerod_2024} find that galaxies become progressively smaller and more compact with increasing redshift. Whilst it had been known for a number of years that galaxy sizes become smaller beyond $z \sim 3$, this was the first time such an analysis had been conducted and this trend seen with JWST data for redshifts beyond this. 

As redshift increases, at around z $> 3$ the size progression starts to taper off, this could be a consequence of less time passing between each redshift value, leading to less time for galaxy sizes to evolve between each higher redshift interval. \citet{varadaraj2024sizesbrightlymanbreakgalaxies} also conducted a similar galaxy size evolution analysis, observing 1668 rest-frame UV-bright Lyman-break galaxies at $3 < z < 5$ with $\log(M_{*}/M_{\odot}) > 9.5$, from the JWST PRIMER survey. The evolution found here is in agreement to those in similar JWST work, which we find to be for our entire sample:

\begin{equation}
    R_{e} = \left( 2.12 \pm 0.28 \right) \left( 1 + z \right) ^{-0.67 \pm 0.06} \ {\rm kpc}.
    \label{eq:power_law}
\end{equation}

\noindent When analysing the galaxies found in our work, we find evidence that this decrease in size at increasing redshift continues into the higher redshift regime. Our average galaxy sizes for each redshift bin can be found in Table \ref{tab:Re_median}. We do find however our galaxies appear to be smaller than what extrapolations predict in previous works, this is likely due to a lack of a mass cut performed at these higher redshifts. The sample used in \citet{ormerod_2024} have masses of $\log(M_{*}/M_{\odot}) > 9.5$, and similarly those used in \citet{varadaraj2024sizesbrightlymanbreakgalaxies} include masses $\log(M_{*}/M_{\odot}) > 9$. At these higher redshifts, no mass cut is performed in order to maximise the size of our sample, as the number of massive galaxies found at these redshifts are much smaller than what is found at lower redshifts.  This is not an ideal comparison, but it is rare to find galaxies at such high masses at high-z, and thus this is a first analysis of the samples we do find.

\begin{table}
    \centering
    \begin{tabular}{c|c|c}
        Redshift & $R_{e}$ (kpc) & S\'ersic Index \\
        \hline
        7 & $0.53^{+0.38}_{-0.18}$ & $0.87^{+1.47}_{-0.73}$\\
        8 & $0.48^{+0.36}_{-0.14}$ & $0.96^{+1.92}_{-0.64}$\\
        9 & $0.44^{+0.27}_{-0.12}$ & $1.21^{+1.40}_{-0.81}$\\
        10.5 & $0.42^{+0.06}_{-0.11}$ & $0.86^{+0.76}_{-0.46}$\\
        12 & $0.37^{+0.16}_{-0.08}$ & $0.84^{+1.60}_{-0.41}$\\
    \end{tabular}
    \caption{Median half-light radii and S\'ersic index values, as measured in the F444W band for each redshift bin. Errors given are 84th and 16th percentiles.}
    \label{tab:Re_median}
\end{table}

Our average sizes are more comparable to the sizes of individual galaxies found in \citet{Yang_2022, Bridge_2019}. Work by \citet{Yang_2022} consists of JWST data of 19 photometrically selected bright galaxies at $z > 7$ within the GLASS ERS program, whose sizes have been measured in their optical rest-frame, finding and demonstrating that NIRCam is capable of imaging half-light radii down to $\sim 100\mathrm{pc}$ at these high redshifts. The galaxies presented in \citet{Bridge_2019}, called ``The Super Eight'', are 8 very luminous high redshift ($7.1 < z < 8.0$) galaxy candidates, using HST and Spitzer data. The sizes of these galaxies are measured using the HST F160W filter, and have masses of the order $\log(M_{*}/M_{\odot}) \sim 10$.

Since our sample population spans a mass range of $\sim 3$ dex, and the observations of the aforementioned proposed break in the size-mass relation \citep{chamba_2024}, we further make a comparison of the redshift evolution of $R_e$ when we split our sample at $M_{\odot} = 4 \times 10^8$ M$_{\odot}$ . In doing so we observe that our more massive galaxy bin ($\log(M_{*}/M_{\odot}) > 8.6$) is larger in size on average, as expected, with galaxies having sizes which are either larger or smaller than the sample as a whole by $\sim 15\%$ in both mass bins. However, we do not find that this significantly alters our results, with our sizes still being smaller than those extrapolated from previous studies. We show these results in Appendix \ref{sec:mass-break}.

\begin{figure*}
    \centering
    \includegraphics[width=1\linewidth]{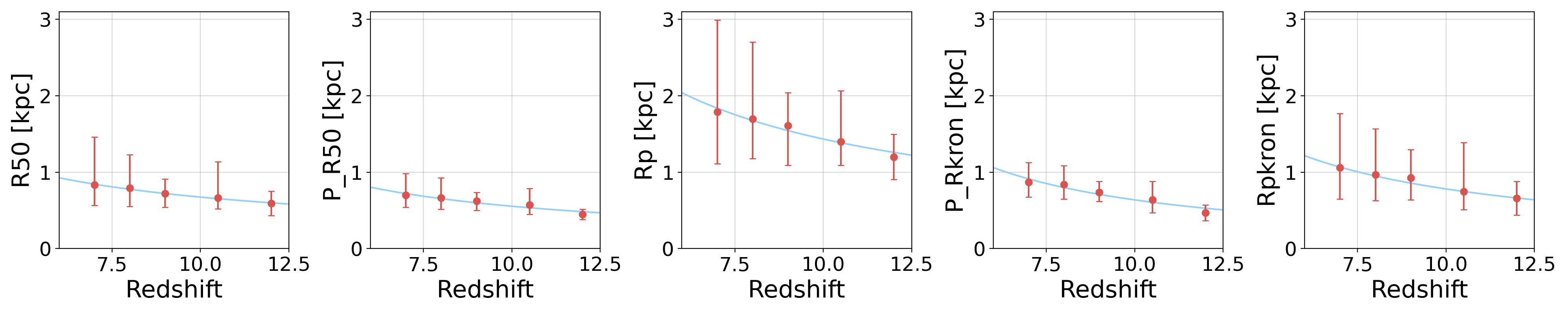}
    \caption{Size evolution as measured by different size parameters from \texttt{Morfometryka}.  This includes the Petrosian radius which measures more of the total size of a galaxy, rather than just its half-light radii.  As can be seen, we find a similar decreasing trend amongst all parameters, giving extra evidence that galaxies have been continuing to grow since $z \sim 12.5$.}
    \label{fig:mfmtk_sizes}
\end{figure*}

We find further evidence of galaxy sizes continuing to decrease at these redshifts by repeating the same analysis with alternative size measurements. Through \texttt{Morfometryka} we take measurements for the Petrosian radius (\textsf{Rp}), the Kron radius measured both through \texttt{photutils} (\textsf{P\_R{kron}}) and by the code itself from within the Petrosian region (\textsf{Rpkron}), as well as the half-light radii from within the Petrosian region (\textsf{R{50}}) and within the Kron region as measured with \texttt{photutils} (\textsf{P\_R{50}}). Each of these parameters follow the same decreasing trend, becoming smaller as redshift increases. These can be seen in Fig. \ref{fig:mfmtk_sizes}.  Since the Petrosian radius is a measure of the `total' size of a galaxy, often measured out to the `edge', this shows that galaxies are growing not just in half-light radii, but also in terms of the total size. 

\begin{figure}[h!]
    \centering
    \includegraphics[width=1\linewidth]{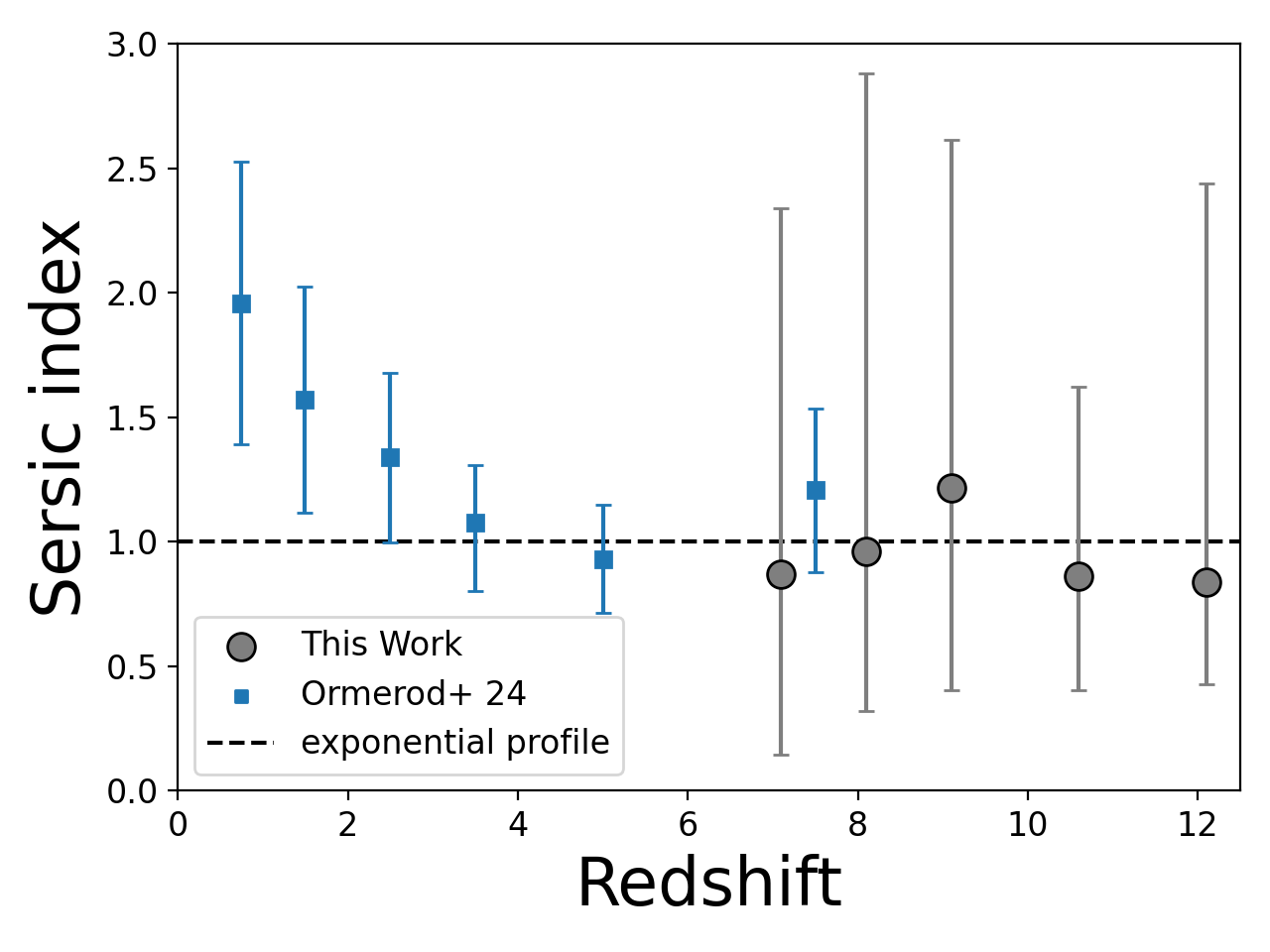}
    \caption{S\'ersic index evolution, as measured through \texttt{GALFIT}. Blue data points show \citet{ormerod_2024} data. Grey points show this work, data points are median values, errors are taken as the 16th and 84th percentiles. Black dashed line at $n = 1$ represents the case of an exponential disc profile.}
    \label{fig:sersic_evolution}
\end{figure}

\citet{ormerod_2024} further finds a trend of decreasing Sersic index $n$ at increasing redshift, before plateauing at $z \sim 5$ to $n \sim 1$ for the highest bins in their analysis, suggesting that there are a higher proportion of disc-type galaxies in the early universe. We plot our average values of n for our redshift bins in Fig. \ref{fig:sersic_evolution}, finding that the average values of n tend to also stay constant at $n \sim 1$ at $z > 6.5$. While we do note this result, it should also be noted that our scatter is much larger than what is seen in \citet{ormerod_2024}.   This is quite interesting and shows that there is a wide diversity of galaxy morphology in the early universe, including for galaxies which have values $n > 3$, similar to giant elliptical galaxies today (Figure~\ref{fig:size_sersic_distribution}). However, these systems are very low mass and therefore may represent a type of compact galaxy which does not exist often in the nearby universe.  In order to better constrain these results, for statistical purposes a larger sample size may be necessary at these higher redshifts, to more reliably and more confidently draw conclusions of how $n$ evolves in the early universe. We further explore how reliably we are measuring S\'ersic indices in Appendix \ref{sec: Simulated Sersic Profiles}. Our average values for $R_e$ and $n$ can be found in Table \ref{tab:Re_median}.

\subsection{C-A Diagram}

\begin{figure*}
    \centering
    \includegraphics[width=1\linewidth]{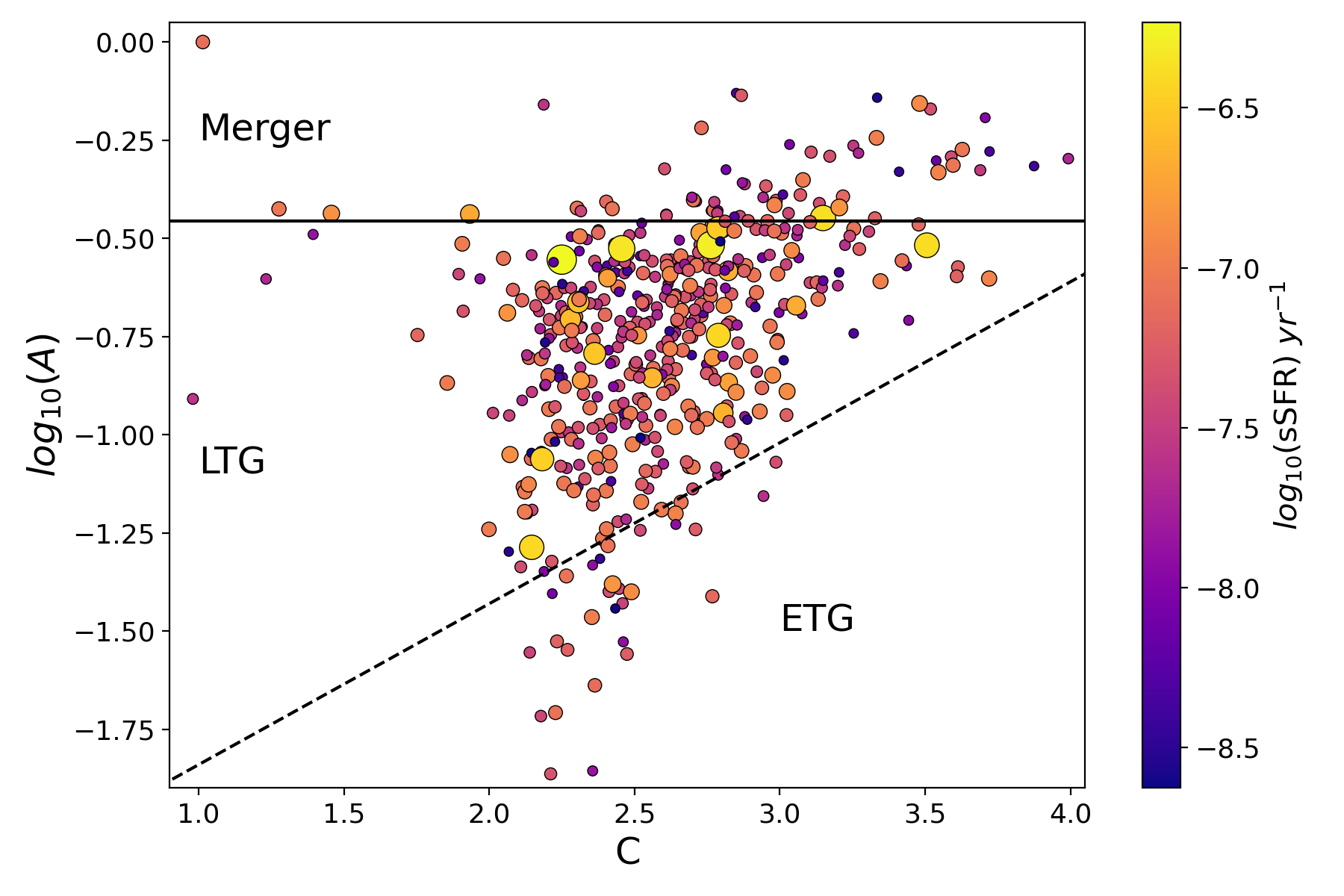}
    \caption{Concentration-Asymmetry diagram for our sample, as measured through \texttt{Morfometryka}. Black solid line represents the merger classification of A $> 0.35$ \citep{Conselice_2003}, whilst the black dashed line represents the late-type vs intermediate types boundary as defined in \citet{Bershady2000}, which is a common method to separate galaxies into late-type galaxies (LTGs) and early-type galaxies (ETGs). Data points are coloured and sized by their specific star formation rates (sSFR), which we see no obvious correlation with.}
    \label{fig:C-A}
\end{figure*}

Another important aspect in analysing the morphology of galaxies, and how they evolve to be the way we see them today, is by analysing the rate at which galaxies merge with each other. One theory as to how galaxies grow in both size and mass, is that galaxies formed much smaller and less massive than what we see today. These smaller galaxies are thought to have merged to create the larger more massive galaxies we see more locally, with larger mergers dictating the shape of the galaxy altogether.

One method to separate galaxies between their morphology, and class those that are merging, is through the non-parametric analysis of the Concentration (C) and Asymmetry (A) parameters. We can classify whether a galaxy in this study is a merger using the widely used asymmetry criteria, as derived through HST work \citep{Conselice_2003}, of:
\begin{equation}
    {A > 0.35} \quad {\&} \quad {A > S.}
    \label{eq:Merger A}
\end{equation}

\noindent We measure the C and A parameters using \texttt{Morfometryka} in the rest-frame optical, for the same galaxies that passed the goodness of fit check for each \texttt{GALFIT} fitting, and plot the parameters against each other in Fig. \ref{fig:C-A}. We find within our sample 63 merger candidates, and derive a merger fraction, defined in equation \ref{eq:merger_frac} as the fraction of classified merger galaxies of the total number of galaxies, where $N_{m}$ is the total number of mergers and $N_{tot}$ is the total number of galaxies.

\begin{equation}
    f_{m} = \left( \frac{N_{m}}{N_{tot}} \right)
    \label{eq:merger_frac}
\end{equation}

\noindent For our sample, we find a merger fraction of $f_m = 0.12 \pm 0.07$, in close agreement with previous JWST work at higher redshifts \citep[][]{duan_2024}.  This agrees very well with how the merger fraction is measured using galaxy pairs, a completely independent approach \citep[e.g.,][]{duan_2024}. In terms of structure \citet{dalmasso2024ratecontributionmergersmass} conducted a similar study, looking at galaxy mergers at $4.0 < z < 9.0$, finding 64 galaxy merger candidates, and a merger fraction of $f_m = 0.11 \pm 0.04$, with no evidence of redshift evolution of $f_m$.

Our work along with the work of \citet{dalmasso2024ratecontributionmergersmass}, extends the previously seen constant trend of merger fractions beyond what was previously seen at $z \leq 6.0$ \citep[e.g.,][]{Conselice2003b, Lin2008, Conselice2009, Bluck2012, Conselice_2014, L_pez_Sanjuan_2009, Jogee2009, Mundy2017, Duncan2019, Kim_2021, Conselice2022}, who all find the merger fraction to oscillate around $f_m \sim 0.10$. From these previous works and our own, there does not seem to be any evidence of the merger fraction having any strong dependencies on redshift after $z\sim 4$, although does increase from $z \sim 0 - 4$.

We further search for any correlation with specific star formation rates (sSFR) with their distribution on the C-A diagram. We colour each object based on its sSFR value in Fig. \ref{fig:C-A}, as measured through \citet{Conselice_2024}, from which we find no correlation to the star formation of our sample with the concentration and asymmetry merger classification. We also find this to be the case for the mass of objects within our sample.  This is further confirmation that structure and star formation rate, as we are measuring both at the current time, do not seem to correlate strongly in the early universe \citep[e.g.,][]{Conselice_2024}.

\subsection{Axis-ratio}

\begin{figure*}
    \centering
    \includegraphics[width=1\linewidth]{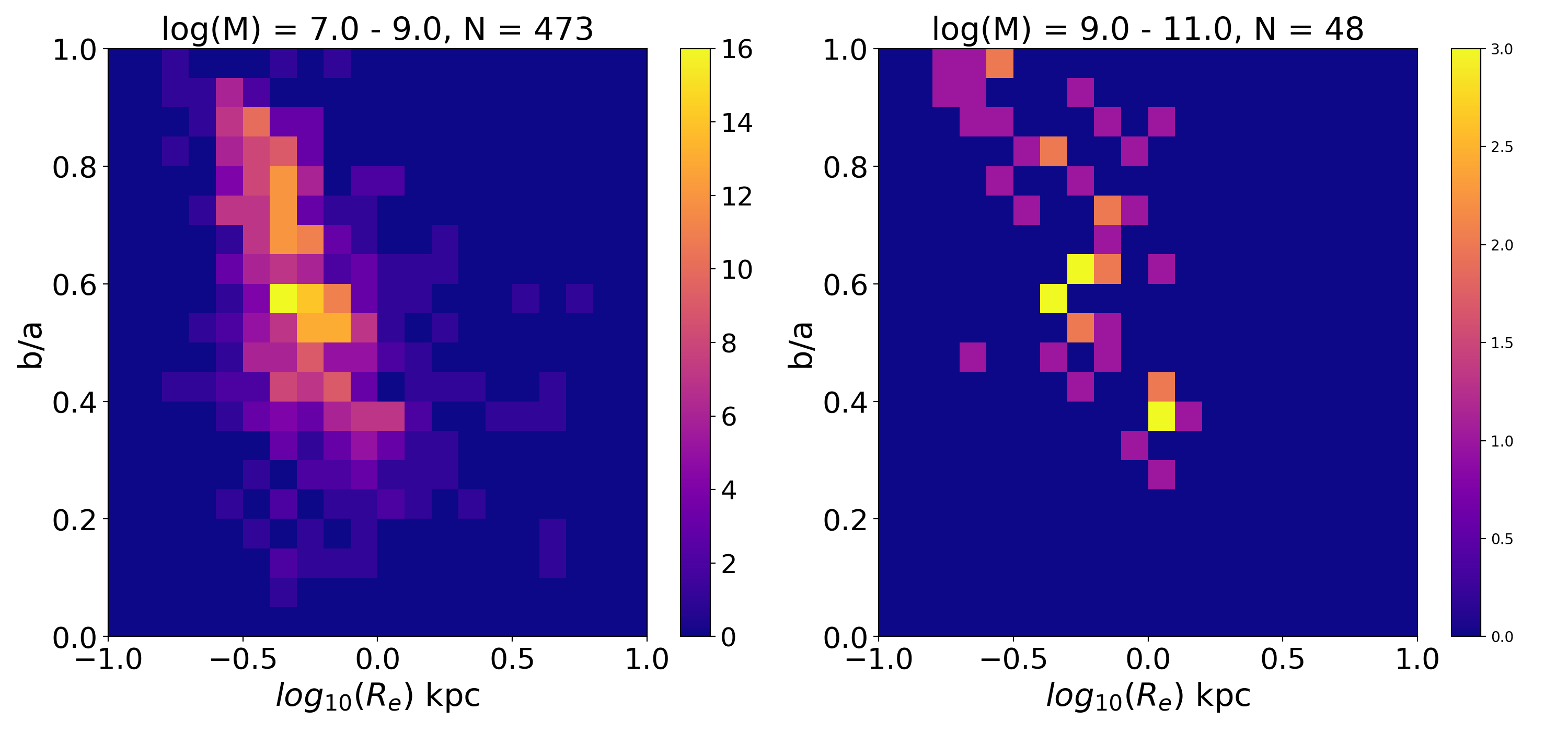}
    \caption{2D histogram of our axis-ratio and galaxy sizes, as measured through \texttt{GALFIT}. Our sample is split between between lower masses (log($M_{*}/M_{\odot}$) $< 9$) and higher masses (log($M_{*}/M_{\odot}$) $ > 9$). Similar to \citet{Pandya_2024}, we find our galaxies trace out a 'banana' shape, with smaller objects appearing rounder.}
    \label{fig:b/a-size}
\end{figure*}

We further analyse the structure of our galaxies by looking at the projection of our sample's axis-ratio with galaxy size. Fig. \ref{fig:b/a-size} shows our projected $(b/a) - \log(R_{e})$ diagram. These plots are split by mass, with the left panel containing galaxies with stellar masses $\log\left( M_{*}/M_{\odot} \right) = 7.0 - 9.0$, and $\log\left( M_{*}/M_{\odot} \right) = 9.0 - 11.0$ in the right panel. Both panels contain galaxies with redshifts $z > 6.5$.

\begin{figure*}
    \centering
    \includegraphics[width=1\linewidth]{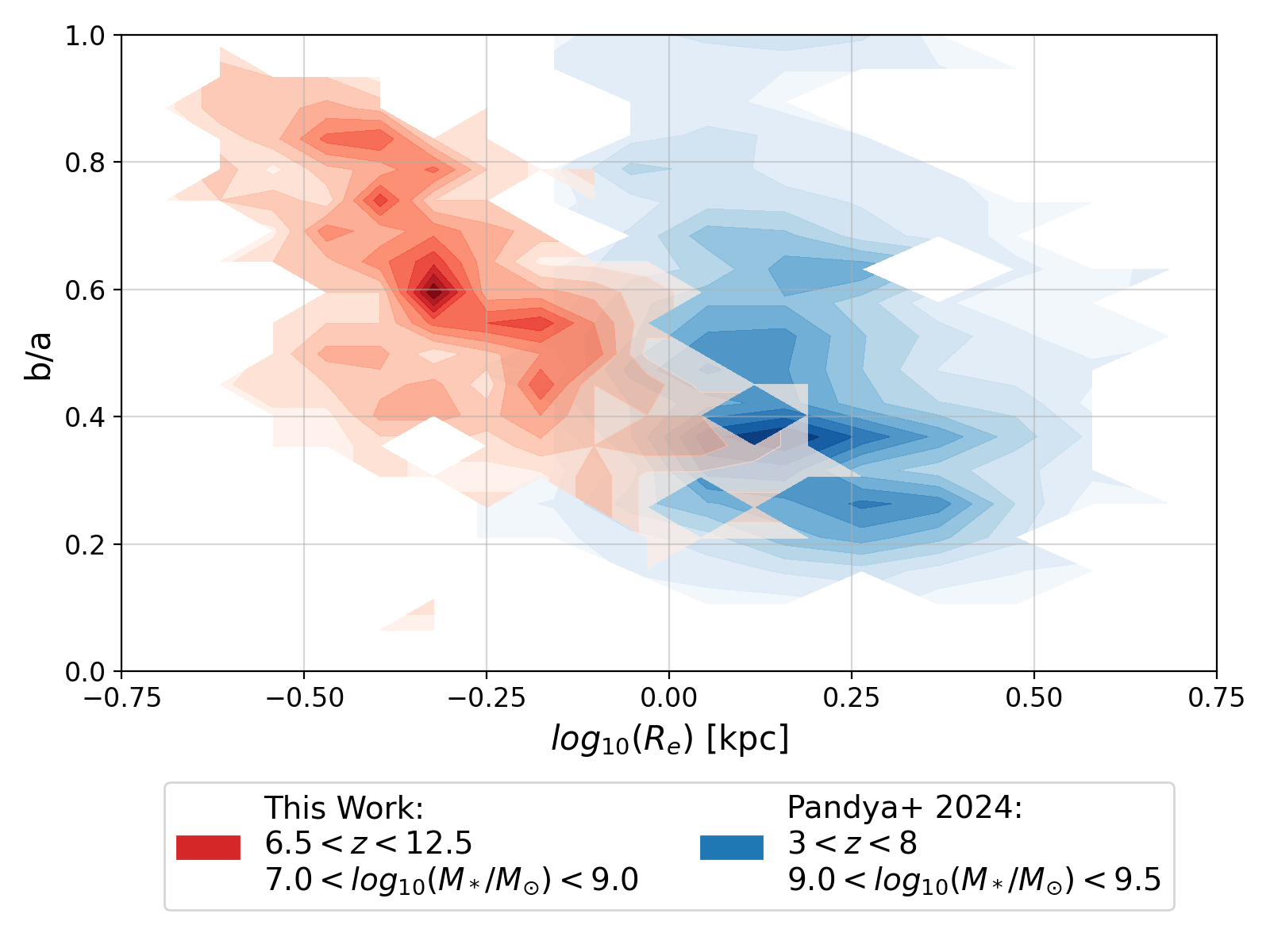}
    \caption{Comparison contour plot of axis-ratio and galaxy sizes, as measured through \texttt{GALFIT}. We compare to the lowest mass, highest redshift bin presented in  \citet{Pandya_2024}, who found a deficit of high b/a objects and an excess of low b/a objects. Our lower mass higher redshift sample, find an opposite trend of high b/a objects and a deficit of low b/a objects. We further find that smaller objects tend to be better fit with much rounder (higher b/a) axis-ratios.}
    \label{fig:pandya_comparison}
\end{figure*}

Similar to the work presented by \citet{Pandya_2024}, focusing specifically on their analysis of CEERS galaxies of mass $\log\left( M_{*}/M_{\odot} \right) = 9.0 - 10.5$ at redshifts $0.5 < z < 8.0$, we find our sample traces out a `banana' shape in the projected axis-ratio size plot, with smaller, more compact galaxies appearing to have a rounder axis-ratio. \citet{Pandya_2024} finds a deficit of rounder objects as mass decreases, leaving an excess of low b/a objects. We compare the distribution of our lower mass bin to the highest redshift, lowest mass bin from work presented by \citet{Pandya_2024}, displayed in Fig. \ref{fig:pandya_comparison}. This work finds a differing trend in comparison, with our lower mass objects being more compact and thus giving us an excess of round objects, with a deficit of low b/a objects.  We examine in the next subsection how reliable these and the other structural results are through the use of simulations, and discuss their implications in \S \ref{sec:discussion}.

\subsection{Comparison with Image Simulations}
\label{sec:simulations}


\begin{figure*}
    \centering
    \includegraphics[width=1\linewidth]{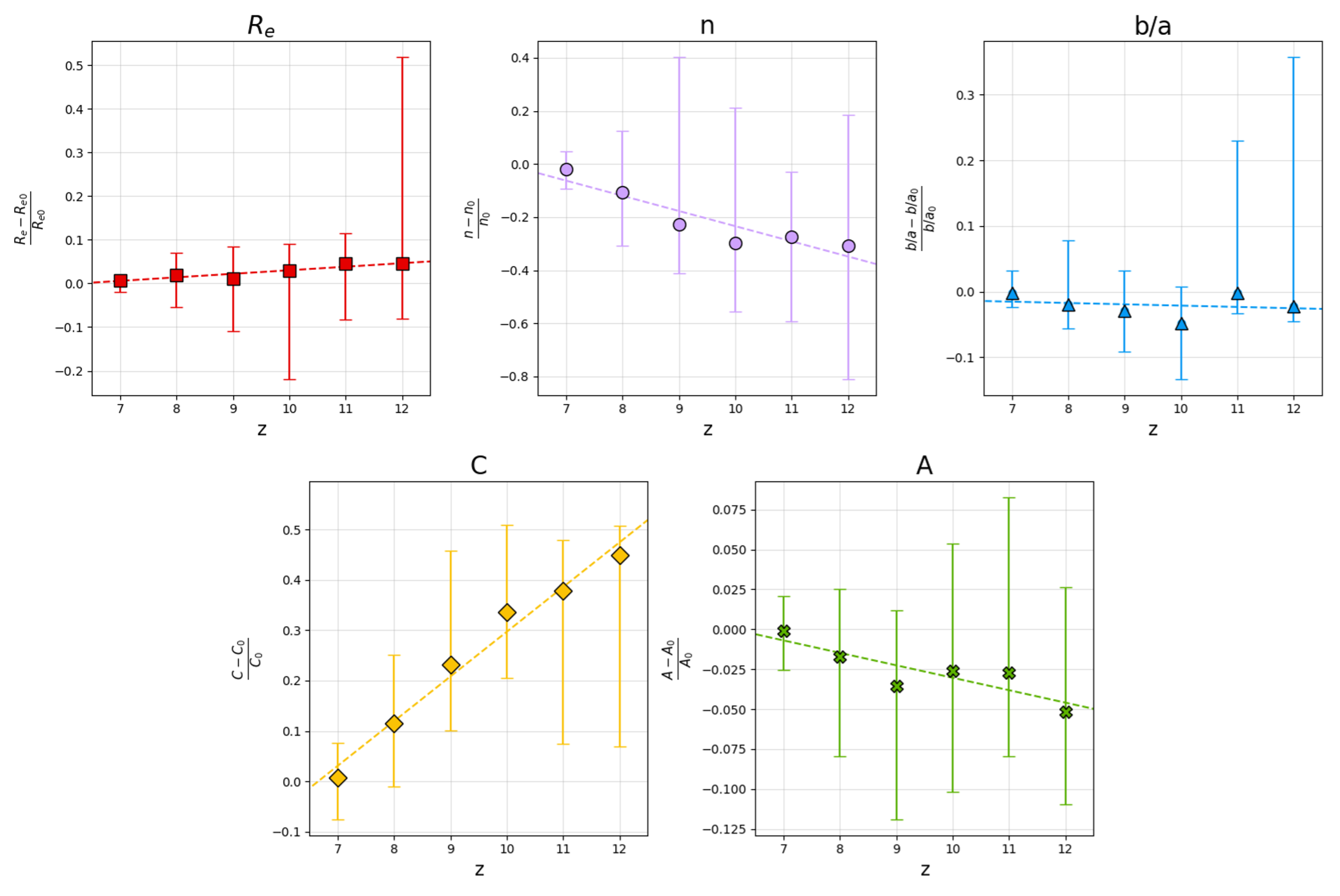}
    \caption{Fractional change of each structural parameter used in this study from redshift simulation images that have been remeasured. Data points are the mean value of each redshift bin, with errors being the standard deviation. Positive values indicate an increase to the parameter value once remeasured at a given simulated redshift. Best fit dashed lines to the data are found through \texttt{emcee}, with gradient values shown in Table \ref{tab:sim_m}.}
    \label{fig:simulated_plots}
\end{figure*}

When observing the structure of galaxies at differing redshifts on this scale, it is important to consider the fact that the surface brightness of galaxies declines as $(1 + z)^{4}$, which can significantly alter the structural appearance of a galaxy as observed through a telescope \citep{Whitney_2020, Whitney_2021}. As a result of this, it is very important to ensure that what we see in our results are as a result of real evolutionary effects, or if they are more likely to be a consequence of the declining nature of the surface brightness of these galaxies causing them to just look smaller. Previous work with HST and JWST up to $z \sim 8$ suggests that we can still accurately measure their structural parameters \citep[e.g.,][]{Whitney_2020, Whitney_2021, ormerod_2024}.

In order to extend our understanding of this issue at the highest redshifts with JWST, we take a sub-sample of 285 of galaxies at redshifts $6.5 < z < 7.5$ from our sample of well fit galaxies, which do not have neighbouring galaxies nearby. Each galaxy has simulated images created at redshift intervals of $\Delta z = 1$ starting at its nearest next redshift integer value, up to a redshift of $z = 12$. These simulated images include all known cosmological effects, to separate real evolutionary effects from redshift effects.

In order to carry out these simulations we use the code \texttt{AREIA}\footnote{https://github.com/astroferreira/areia}. Here we give a brief description of the code, but for more details on how the code works please see \citet{2021Tolhill} and \citet{Whitney_2021}.

The source galaxy is first extracted from the original cutout, which is done through constructing a segmentation map which is handled through the code \texttt{GALCLEAN}\footnote{https://github.com/astroferreira/galclean}. From here, the image is geometrically re-binned, taking it from the redshift of the source, to the desired targeted redshift, all while preserving the sources flux. Doing so ensures that sources simulated to higher redshifts have the correct geometric scaling due to the adopted cosmology. Once done, the re-binned image has its flux scaled to follow cosmological dimming effects up to the desired redshift, shot noise is also sampled from each source's new light distribution and convolved with the JWST F444W PSF. Finally the source, now at the desired redshift, is placed onto a random real background image taken from the same field that the source is originally from. Doing this for each of our sub sample galaxies gives us a total of 1294 simulated images.

To measure the structural parameters of our simulated images, we follow the same methodology as detailed in section \ref{sec:GALFIT}, and follow the same selection criteria detailed in section \ref{sec:GALFIT criteria}, which our original sample are subjected to, in order to ensure our measurements are robust. Each simulated image has its signal-to-noise remeasured by using the same local depths, and taking photometry within the same 0.32 arcsecond aperture of the simulated image. From this we only measure those simulated images which have SNR $> 10\sigma$. 

To compare how our redshift simulations affect their measured structural parameters, we focus on the fractional change between the parameter at its photometric redshift and the parameter at a given simulated redshift, of the form $\frac{\left(x - x_0\right)}{x_0}$ where $x$ is the parameter measured at a simulated redshift, and $x_0$ is the parameter value at the objects photometric redshift. In order to avoid the possibility of \texttt{AREIA} adding any artifacts that may affect the measurements between a real image and the simulated images, we also simulate the source at its initial photometric redshift, and use the measurements taken from these images to compare to the measurements from galaxies simulated at other redshifts.

We plot the average fractional changes for each parameter used in this work, along with their uncertainty in Fig. \ref{fig:simulated_plots}, with positive fractional changes indicating the parameter value increasing when simulated to a higher redshift. We further obtain lines of best fit to each parameter using the \texttt{emcee} python package \citep{Foreman_Mackey_2013}. We find a linear fit best fits our simulated data for each parameter, with gradients and gradient errors shown in Table \ref{tab:sim_m}.

\begin{table}
    \centering
    \begin{tabular}{cc}
        Parameter & Gradient\\
        \hline
        Half-light Radius ($R_{e}$ [kpc]) & $0.008 \pm 0.002$ \\
        S\'ersic index (n) & $-0.057 \pm 0.012$ \\
        Axis-Ratio (b/a) & $-0.002 \pm 0.005$\\
        Asymmetry (A) & $-0.008 \pm 0.007$ \\
        Concentration (C) & $0.089 \pm 0.007$\\
        \hline
    \end{tabular}
    \caption{Results from best fit lines for fractional changes of each structural parameter from redshift simulations, as found through the emcee python package. We find that our gradients are relatively flat, in particular for sizes, axis-ratio and asymmetry, with a slight negative slope for S\'ersic index and a steeper increase in concentration.}
    \label{tab:sim_m}
\end{table}


We find for our sub-sample that average galaxy sizes, axis-ratio measurements and asymmetry values do not differ significantly when simulated at higher redshifts, showing a relatively flat slope. For S\'ersic index values we find a slightly steeper negative slope, decreasing on average by $\sim 30\%$ over our redshift range. We do find that concentration measurements differ the most significantly over our redshift range, increasing on average by $\sim 50\%$. While these two parameters vary more significantly, we do not use them in this analysis to make any conclusions, but it should be noted for any future work.

These results indicate that our observed trend of decreasing galaxy sizes with redshift for our real objects, as well as observed higher axis-ratio values, are due to real evolutionary effects and not redshift effects, with simulated galaxy images having very similar size measurements at all redshifts, showing that the trends seen here are due to a change in galactic properties at increasing redshift. Moreover we see little change in asymmetry values which indicates that merger classifications are also not impacted by affects due to redshift.

\section{Discussion}
\label{sec:discussion}

The results presented in this work reveal a range of observational trends in galaxy structure out to a redshift which has not yet been studied before in great detail. One of our results shows us that galaxies continue to get progressively smaller as redshift increases, even up to $z \sim 12$. The evolutionary decrease in galaxy sizes has long been observed, through HST \citep[e.g.][]{Fergurson_2004, Trujillo_2007, Hathi_2008, Buitrago2008, VanDockum2010, vanderWel_2012} and more recently through JWST \citep[e.g.][]{ormerod_2024, varadaraj2024sizesbrightlymanbreakgalaxies}, but now with the development of deeper JWST programs and the creation of one of the largest ultra high-redshift catalogues, EPOCHS v1, we are able to extend these observations to some of the youngest, most distant galaxies ever observed. The continuation of this trend reveals to us that the growth of galaxy sizes is ongoing, even in a galaxy's youngest phase. While this growth does not appear at first glance to be as dramatic as what is seen at lower redshift, the time difference between redshift intervals is much smaller at higher redshifts than at lower redshifts, so this growth is occurring in a shorter period of time.  We also find that this grown is not only in the effective radius, but also in the total or Petrosian radius, meaning that galaxies are growing in size throughout their structure.

In order to maximise the number of galaxies analysed, we do not perform a mass cut in this study, with most galaxies in this epoch of the universe having smaller masses than the cuts made in previous works, this is perhaps one reason for our average galaxy sizes being smaller than what has been predicted in previous works \citep{VanDockum2010, Costantin2023, ormerod_2024, varadaraj2024sizesbrightlymanbreakgalaxies}.

To better observe the evolutionary tracks of more massive galaxies, we will require more area to be imaged through JWST and Euclid to increase the number of high mass galaxies out to the highest redshifts. We do however still see similar trends in our extrapolated power law towards lower redshifts, than what is observed at these redshifts, with the growth of galaxies becoming more dramatic at lower redshifts.

We also find within our sample, on an individual basis, a variety of sizes to our galaxies, including a number of very compact galaxies, with our smallest being just $R_{e} \sim 173$pc. Within our sample we have a total of 57 galaxies with a measured half-light radius $< 300$pc and 51 being $> 1$kpc. The variety of sizes within our sample signifies a complex diversity to the evolutionary paths taken even in the early universe. Recent work with JWST has found similar size distributions at such high redshift \citep{Yang_2022, Ono_2023}, with \citet{Ono_2023} finding a galaxy as compact as $R_{e} = 39 \pm 11$pc.

It should be taken into account that with our sample spanning a mass range of $\sim 3$ dex, it is likely that we are observing galaxies that are undergoing different growth mechanisms. It has been seen that galaxies may experience a break at $4 \times 10^8 M\odot$, which differentiates galaxies into two mass regimes, where lower mass galaxies regulate their sizes more rapidly due to feedback and environmental processes \citep{chamba_2024}. We compare this to our sample, splitting galaxies into a high mass and low mass regime separated at $4 \times 10^8 M\odot$. In finding a steeper size-mass relation for lower mass galaxies within our sample, we provide further evidence to support a potential mass break at $4 \times 10^8 M_{\odot}$.

It is thought one of the major contributors to the growth of galaxies is due to galaxy mergers, likely minor mergers. It has been found in previous JWST studies that mergers cause galaxies to increase in stellar mass by up to a factor of $\sim 2.77$ on average between $5 < z < 10.5$, with merger rates peaking and stabilising at $z > 6$ \citep{duan_2024}. Minor mergers are very difficult to measure at our high redshifts, but the numbers at low redshifts are enough to account for the evolution in size at $z < 3$ \citep[e.g.,][]{Bluck2012}.   It has also been found that merger fractions remains on average fairly constant at $z < 9$, with values oscillating around $f_{m} \sim 0.10$ \citep{Lin2008, Conselice2009, L_pez_Sanjuan_2009,
Mundy2017, Duncan2019, Kim_2021, Conselice2022, dalmasso2024ratecontributionmergersmass}.   

In this work, we consolidate this consistency of merger fractions out to the highest redshifts, finding a merger fraction of $f_m = 0.12 \pm 0.07$ for our sample as a whole using the CAS method, agreeing with prior work at lower redshifts. These results signify that mergers play a key role in galaxy growth throughout cosmic time, linking to the continuous size evolution seen in this work. A constant merger fraction also points to hierarchical structure formation (i.e. galaxy growth through the mergers of smaller structures) being well underway even in the earliest stages on galaxy evolution and significant galaxy interactions were already taking place.

Another finding in this work is an excess of round objects (low $b/a$) for lower mass galaxies within our redshift range, and a deficit of low $b/a$ objects, as seen in Fig \ref{fig:b/a-size}. We further find that galaxies that are smaller and more compact appear rounder, tracing the same `banana' shape as found in \citet{Pandya_2024}. What differs here is that \citet{Pandya_2024} found there to be an excess of low $b/a$ objects, and a deficit of round objects for their lower mass ($9 < \log(M_* / M_{\odot}) < 9.5$), high redshift ($3 < z < 8$) galaxies. One potential reason for this is that we may be seeing rounder objects as our galaxies are younger and may be incomplete to any potential other galaxies that are not currently bursting with star formation, or are post-starburst. A further reason we could be finding rounder objects is due to these higher redshift objects having a more diffuse component that is too faint for JWST to detect, causing us to just be viewing a brighter core. This would lead to increasing axis ratio measurements, indicating rounder objects. Future work could test this hypothesis by stacking these objects to see if potential diffuse components emerge. Another possible reason for us finding rounder objects, could be due to our objects being much smaller in size, with semi-major and semi-minor radial components each being just a few pixels across.  At these scales, a S\'ersic model that differs to the real object by just a few pixels can lead to much more significant variations in the measured axis-ratio, potentially leading to higher values.   However, our simulations of galaxies, placed at higher redshifts, suggests that this is not a major effect.   We further explore this possible effect on simulated single S\'ersic surface brightness profiles in Appendix \ref{sec: Simulated Sersic Profiles}, finding values to vary much more for smaller objects, but not significantly enough that we would expect to see a bias towards rounder objects because of this.

\section{Conclusions}
\label{sec:conclusions}

We present a structural analysis of 520 galaxies at $z = 6.5 - 12.5$ from the EPOCHS v1 catalog \citep[][]{Conselice_2024}, each with a measured SNR $> 10\sigma$, taken from a combination of public ERS and GO data (CEERS, JADES, GLASS, NGDEEP and SMACS-0723) and from the GTO PEARLS program (NEP-TDF, El Gordo, Clio and MACS-0416). To each galaxy we fit a single S\'ersic surface profile model, in order to extract and analyse their parametric structural parameters (S\'ersic index (n) and half-light radius ($R_{e}$)), and determine how these parameters evolve through time. We perform further morphological measures to extract their non-parametric structural parameters (Asymmetry (A) and concentration (C)), to determine the fraction of galaxies which can be classified as mergers based on the widely used CAS criteria \citep{Conselice_2003}. We also use these parameters to probe how the size-mass relation evolves at high redshift, and compare our results to simulations, as well as look at the axis-ratio distribution of our sample.

In order to carry out our analysis, we measure each object's parametric structural parameters using a custom built \texttt{GALFIT} pipeline, in which we take measurements from a single S\'ersic fit, modelling nearby neighbours when appropriate, and our non-parametric structural parameters are measured through \texttt{Morfometryka}. 

We verify that our results are robust against redshift effects, as we show through \texttt{AREIA} redshift simulations, finding the difference in each parameter at each redshift interval between the simulated redshift and the objects photometric redshift.

From our analysis, our main results are as follows:

\begin{itemize}
    \item Our observed galaxy sizes continue to show a slightly decreasing trend with redshift, as has been seen in many previous works with both HST and JWST. We find our galaxies become smaller following the power law $R_{e} = 2.12 \pm 0.28 \left( 1 + z \right) ^{-0.67 \pm 0.06}$. We find our galaxies are on average more compact than what is predicted from extrapolated power laws in literature, likely due to having not performed a mass cut. This is done to maximise the number of galaxies used in our analyses, since more massive objects in the early universe are much less common.
    \item We further observe an evolution to the size-mass galaxy relation, with galaxies of a given mass growing in size as time passes, suggesting galaxies form with denser structures in the early universe. We observe a galaxy of given mass $\log(M_{*}/M_{\odot})$ = 8 to grow in size by $\sim 53\%$ in $\sim 400$ Myr. We find our observations follow what is seen in simulations at slightly lower redshifts and more massive galaxies.
    \item We find galaxy merger fractions for our whole sample in the early universe to have a value of $f_{m} = 0.12 \pm 0.07$. This is consistent to that found in previous JWST and HST studies, which has seen the value for merger fractions oscillating around $f_{m} \sim 0.1$ independent of redshift at $z < 9$. This relatively constant merger fraction remaining out to $z \sim 12$ points to significant galaxy interaction already being established in the early universe, causing the growth in size and mass we see through cosmic time.
    \item We find an excess of round (high b/a) objects in the early universe, and a deficit of low b/a objects, with smaller sized objects being rounder, tracing out a similar `banana' shape as seen in \citet{Pandya_2024}.
    \item We run redshift simulations on our objects, to see if our observed trends are due to redshift effects or are real evolutionary effects. From our remeasured structural parameters, we find flat trends for our half-light radii, axis-ratio and asymmetry measurements, substantiating that any observed trends with these parameters are due to real evolutionary effects. We do see more significant changes for parameters not used to determine trends in this work, with a small decrease in S\'ersic index values and a more significant increase to concentration measurements.
\end{itemize}

Overall in this paper, we have shown that the structure of galaxies continues to evolve from the highest redshifts, with galaxies becoming more compact, with denser structures the further back we look. We also find that merger fractions continue to be relatively redshift independent into the highest redshifts, giving us an insight into the amount of galaxy interactions ongoing in the early universe.

Our observations into the high redshift universe, and in particular the structures of galaxies from this epoch, are still only just beginning with JWST. As the JWST mission evolves, we will have access to even deeper data, from a growing total observation area, which will allow these subtle changes to structural features and sizes to become more apparent, allowing us to further develop our understanding of galaxy evolution from the earliest moments in cosmic time.

\section*{Acknowledgements}

\vspace{10pt}

LW, CC, NA, DA, TH, QL, JT, VR, QD, HH, CG acknowledge support from the ERC Advanced Investigator Grant EPOCHS (788113), as well as two studentships from the STFC and a studentship from the Faculty of Science and Engineering at the University of Manchester. RAW, SHC, and RAJ acknowledge support from NASA JWST Interdisciplinary Scientist grants NAG5-12460, NNX14AN10G and 80NSSC18K0200 from GSFC. LF acknowledges financial support from Coordenação de Aperfeiçoamento de Pessoal de Nível Superior - Brazil (CAPES) in the form of a PhD studentship.

This work is based on observations made with the NASA/ESA \textit{Hubble Space Telescope} (HST) and NASA/ESA/CSA \textit{James Webb Space Telescope} (JWST) obtained from the \texttt{Mikulski Archive for Space Telescopes} (\texttt{MAST}) at the \textit{Space Telescope Science Institute} (STScI), which is operated by the Association of Universities for Research in Astronomy, Inc., under NASA contract NAS 5-03127 for JWST, and NAS 5–26555 for HST. The authors thank all involved with the construction and operation of JWST, without whom this work would not be possible. We also thank the PI's and teams who designed and executed the ERS, GTO and GO programs used within this work, including PEARLS (1176, 2738), SMACS-0723 (2737),  GLASS (1324), CEERS (1345), JADES (1180, 1210, 1895, 1963)
and NGDEEP (2079).

This work makes use of {\tt astropy} \citep{astropy_2013,astropy_2018,astropy_2022}, {\tt matplotlib} \citep{hunter_2007}, {\tt reproject}, {\tt DrizzlePac} \citep{hoffmann_2021}, {\tt SciPy} \citep{SciPy-NMeth_2020} and {\tt photutils} \citep{larry_bradley_2024_12585239}.



\bibliography{high_z_structure}{}
\bibliographystyle{aasjournal}

\appendix

\counterwithin{figure}{section}

\section{Fitting with a constant S\'ersic index}
\label{sec:holding n}
For our sample, as stated in \S \ref{sec:GALFIT criteria}, we attempt to better model the 119 objects that did not meet the selection criteria by holding constant the S\'ersic index parameter during the \texttt{GALFIT} fitting process. In our work we opt to hold the S\'ersic value at the lower limit of 0.05, after testing holding the S\'ersic index the different benchmark values of $n = 1$ and $n = 4$, typically associated with spiral and elliptical galaxies respectively. 

\begin{figure*}
    \centering
    \includegraphics[width=1\linewidth]{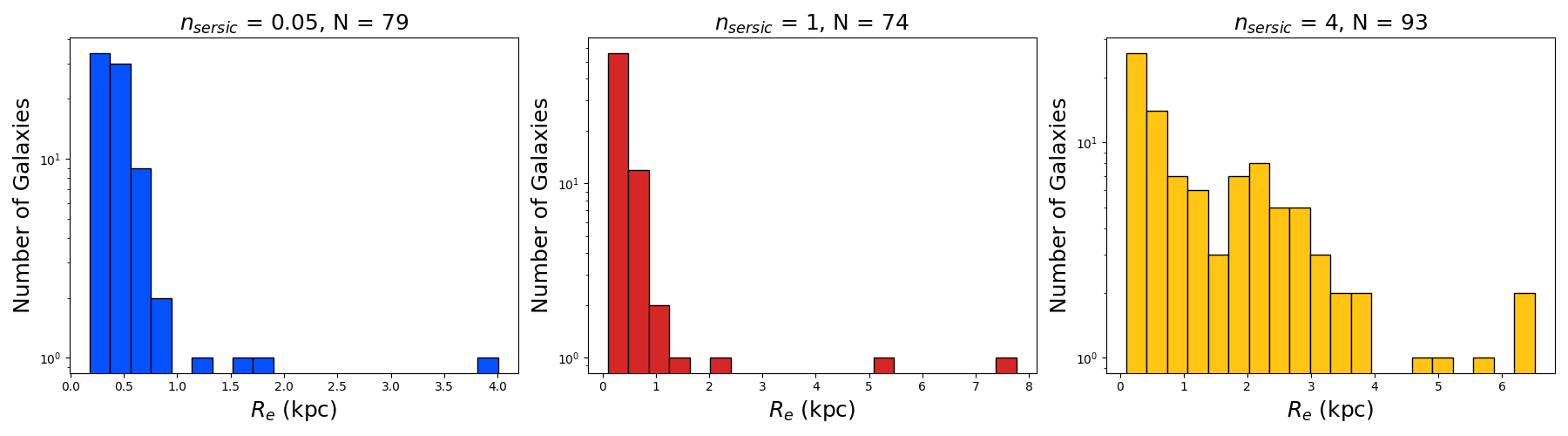}
    \caption{Distribution of galaxy half-light radii after holding the S\'ersic value at different values. Left panel: Size distribution for n = 0.05. Middle panel: size distribution for n = 1. Right panel: size distribution for n = 4.}
    \label{fig:held_n}
\end{figure*}

\begin{figure*}
    \centering
    \includegraphics[width=1\linewidth]{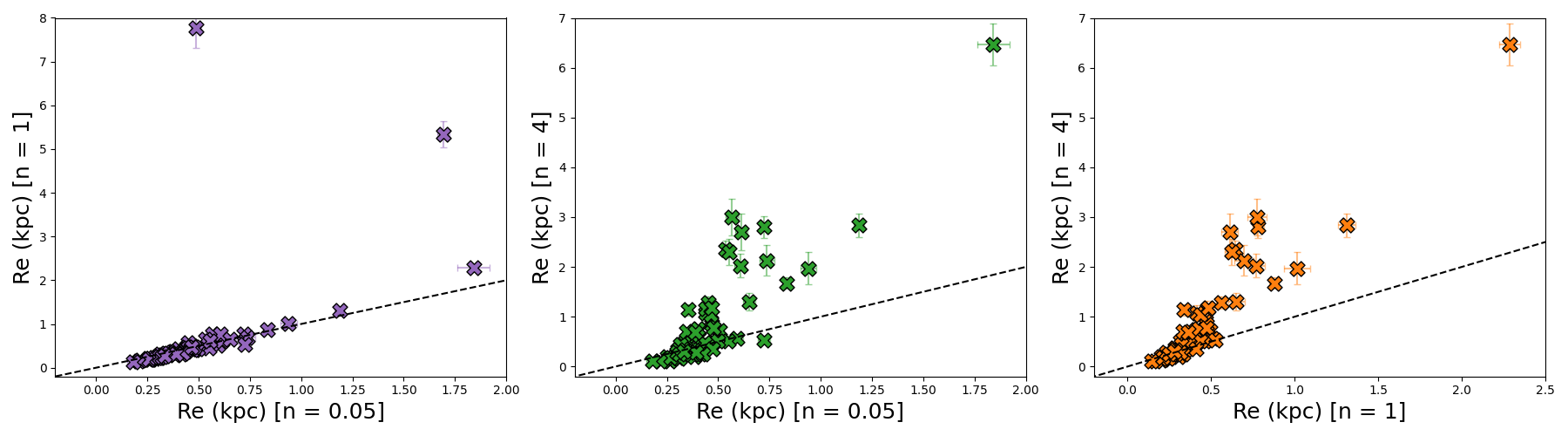}
    \caption{Comparison of half-light radii for different  S\'ersic values held constant. In each plot we show the objects whose models passed the selection criteria for both values of n. The black dashed line shows where the two runs are equal. Left panel: Size values with a held S\'ersic value of 0.05 vs 4. Middle panel: Size values of a held S\'ersic values of $n=0.05$ vs $n=4$. Right panel: Size values of a held S\'ersic value of 1 vs 4.}
    \label{fig:held_n_scatter}
\end{figure*}

Fig. \ref{fig:held_n} shows the different size distributions for models that meet our selection criteria outlined in \S \ref{sec:GALFIT criteria} for different constant values of the S\'ersic index. Additionally Fig. \ref{fig:held_n_scatter} compares galaxy half-light radii values between different values at which the S\'ersic index is held.

We choose to use the half-light radii values which are output from the models where the S\'ersic value is held at $n = 0.05$. We find our best fit \texttt{GALFIT} models where $n = 0.05$ and $n = 1$ to be fairly equal for the vast majority of our objects, with only two objects having significantly larger sizes than when $n = 1$. Comparing these two sub-samples to the instance where $n = 4$, galaxy sizes are modelled to be much larger, with some values being unphysically large for this epoch, there are also very few galaxies at these higher redshifts that have a S\'ersic index value of $n = 4$. Due to this we opt to discard the option of using a held S\'ersic index of $n = 4$. We find through free fits of this index that $n=4$ is extremely rare at higher redshifts, thus backing up our conclusion.

\section{Changes in Wavelength}
In this work we observe the structure of objects using the F444W LW NIRCam filter, in order to best measure the rest-frame optical structures of galaxies at these high redshifts. Here we make further observations with the F410M NIRCam filter to see if any of the measured parameters and observed trends we find in our work, are altered with a change in observed wavelength to $\sim 4.1\mu m$.

In order for this comparison to be robust, we repeat the selection criteria as outlined in \S \ref{sec:morphological fitting} for our objects, this time in the F410M NIRCam filter, first requiring each object to have a SNR $> 10\sigma$ in the F410M band. Following this criteria, we start with a sample of 299 galaxies through the F410M filter at $> 10\sigma$, for which we fit a single S\'ersic surface brightness profile to. Upon reviewing their best fit models to our selection criteria and the RFF of the model, we cut 64 objects, leaving us with a final sample of 235 well fit galaxies in the F410M NIRCam band.

\begin{figure}
    \centering
    \includegraphics[width=\linewidth]{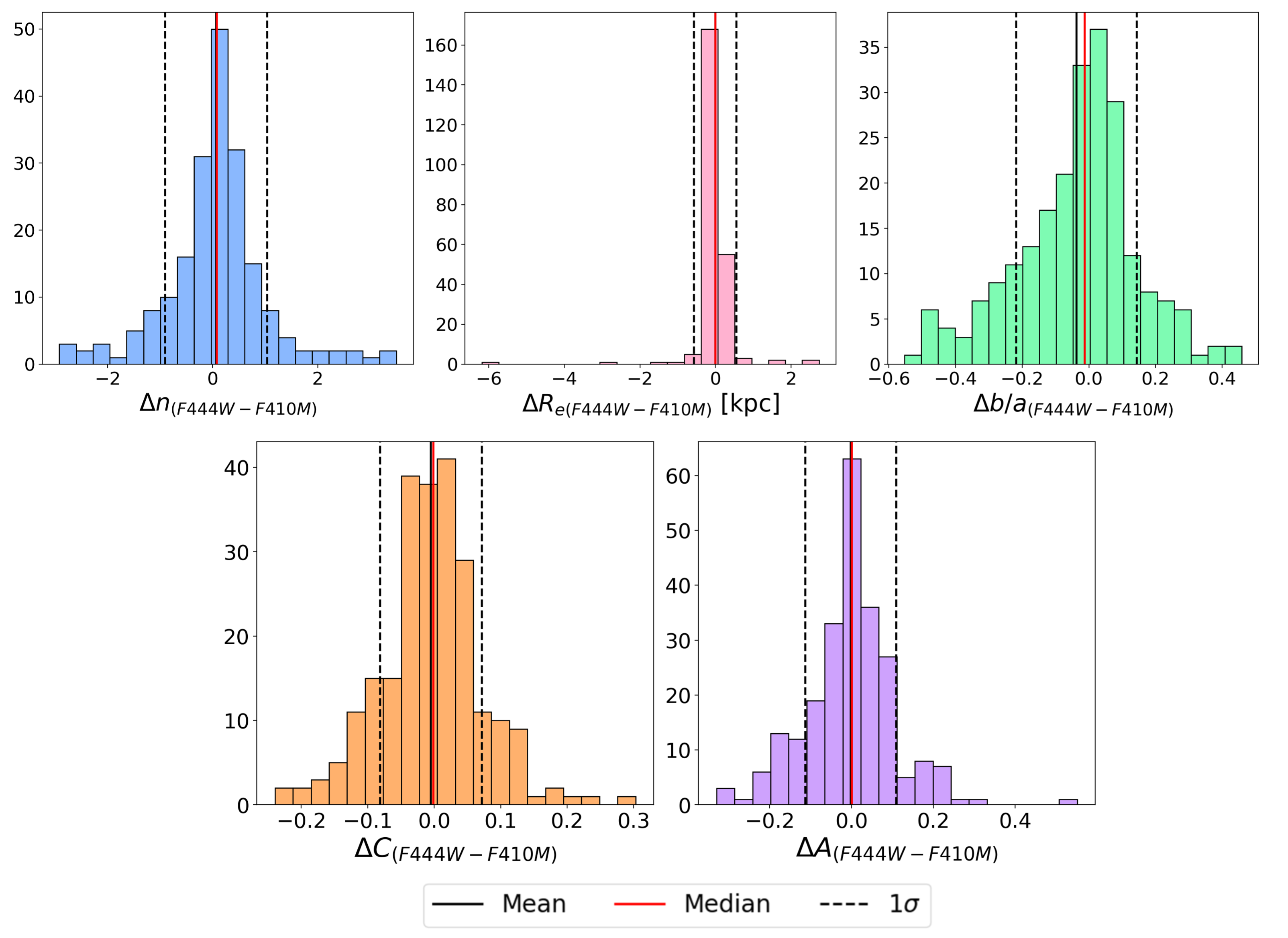}
    \caption{Distribution of changes to structural parameters from F444W to F410M ($\Delta x = x_{F444W} - x_{F410M}$). Solid black line represents the mean of the distribution. Solid red line represents the median of the distribution. Dashed black line represents a standard deviation from the mean.}
    \label{fig:F410M_hist}
\end{figure}

\begin{figure}
    \centering
    \includegraphics[width=\linewidth]{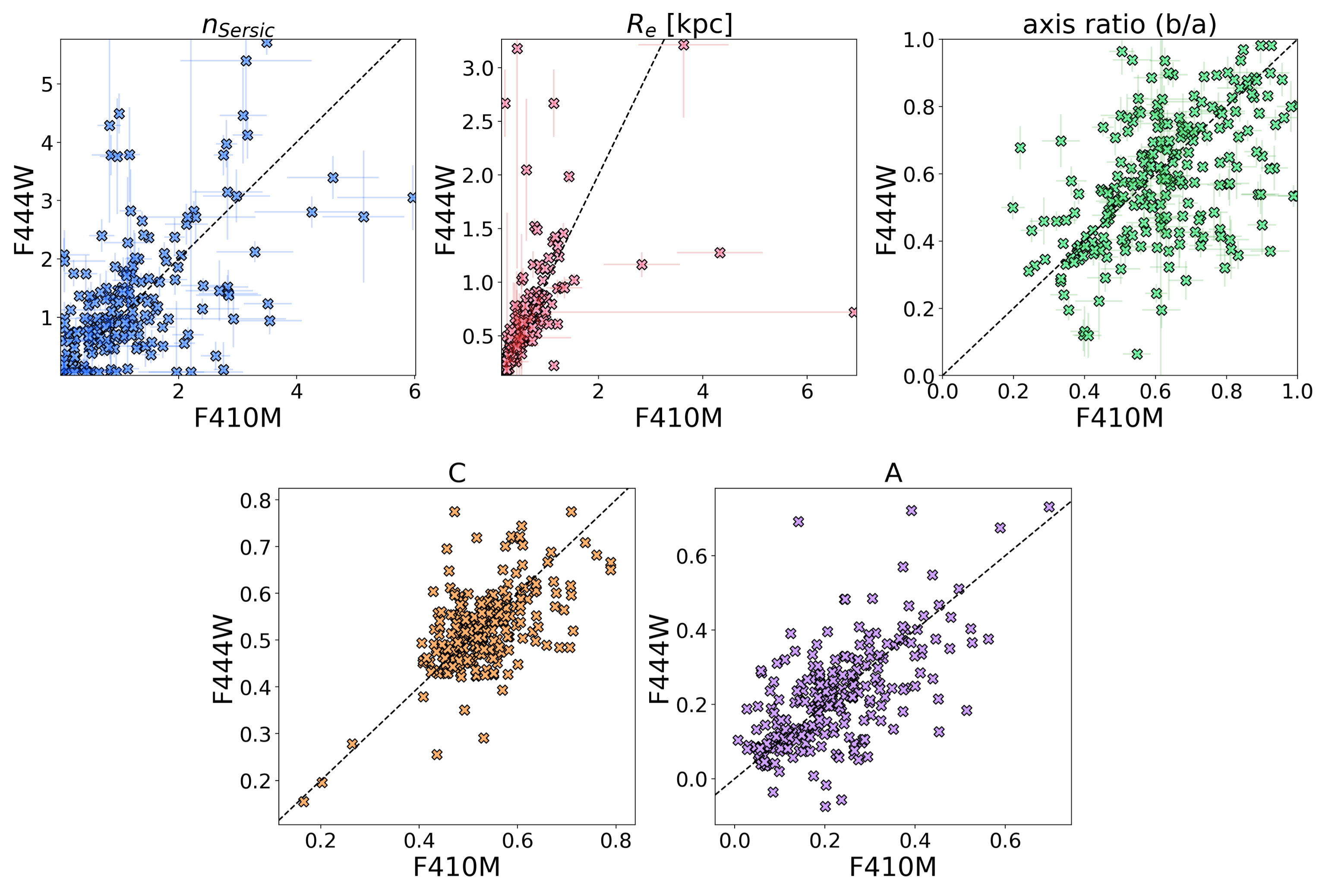}
    \caption{Comparison of parameter measurements at different wavelengths, through the F444W and F410M bands. Black dashed line represents where the two are equal.}
    \label{fig:F410M_scatter}
\end{figure}

Fig. \ref{fig:F410M_hist} shows the distribution of the difference in values of each parameter ($\Delta x$) between F444W and F410M for their respective best fit models. While Fig. \ref{fig:F410M_scatter} compares the directly measured values for each parameter in the F444W and F410M bands. For these comparisons, we only examine S\'ersic values that were not held constant during the fitting process.

From these comparisons, while we observe that on a case by case basis some parameters may vary a notable amount, we see that the change in value to all parameters on average is minimal, with all averages showing a change in parameter value of $\sim 0$. This is reflected in the observed trends reported upon in this work, where we find similar trends using data from F410M. We find a similar decrease to average galaxy sizes with increasing redshift, we see the same evolution to the size-mass relation where galaxy size decreases for a given mass at increasing redshift and we see an excess of round (high b/a) objects with a deficit of low b/a objects.

To quantify how significantly the measurements of each parameter changes within these observed wavelengths, we calculate the Normalised Median Absolute Deviation (NMAD), defined in equation \ref{eq:NMAD}, which measures the spread of the parameter measurements in F410M around the measurements in the F444W filter \citep{hoaglin_1983}. Table \ref{tab:NMAD} shows the NMAD values obtained for each parameter.

\begin{equation}
    NMAD = 1.4826 \times {\rm Median} \left| \left( x_{F444W} - x_{F410M} \right) \right|
    \label{eq:NMAD}
\end{equation}

\noindent We find a good agreement between most parameters at the two measured wavelengths, with the exception of the S\'ersic index, where we obtain a high NMAD value. As mentioned in \S \ref{sec:re evolution}, we are not able to draw reliable conclusions on how the S\'ersic index evolves due to large uncertainties and scatter from F444W measurements, so while we do not find much agreement between the measurements for n in these two wavelengths, this does not massively affect the outcome of this work, and could instead point to the S\'ersic index not being a reliable measure at these high redshifts.

\begin{table}[h!]
    \centering
    \begin{tabular}{c|c}
        Parameter & NMAD\\
        \hline
         S\'ersic index (n) & 0.592\\
         Half-light radius ($R_{e}$) & 0.089\\
         Axis-ratio (b/a) & 0.144\\
         Concentration (C) & 0.06\\
         Asymmetry (A) & 0.077\\
    \end{tabular}
    \caption{NMAD values comparing parameter values at different wavelengths, as measured through F444W and F410M NIRCam filters.}
    \label{tab:NMAD}
\end{table}

\section{Effects of PSF Variation}
\label{sec:PSF_comparison}
\begin{figure}
    \centering
    \includegraphics[width=1\linewidth]{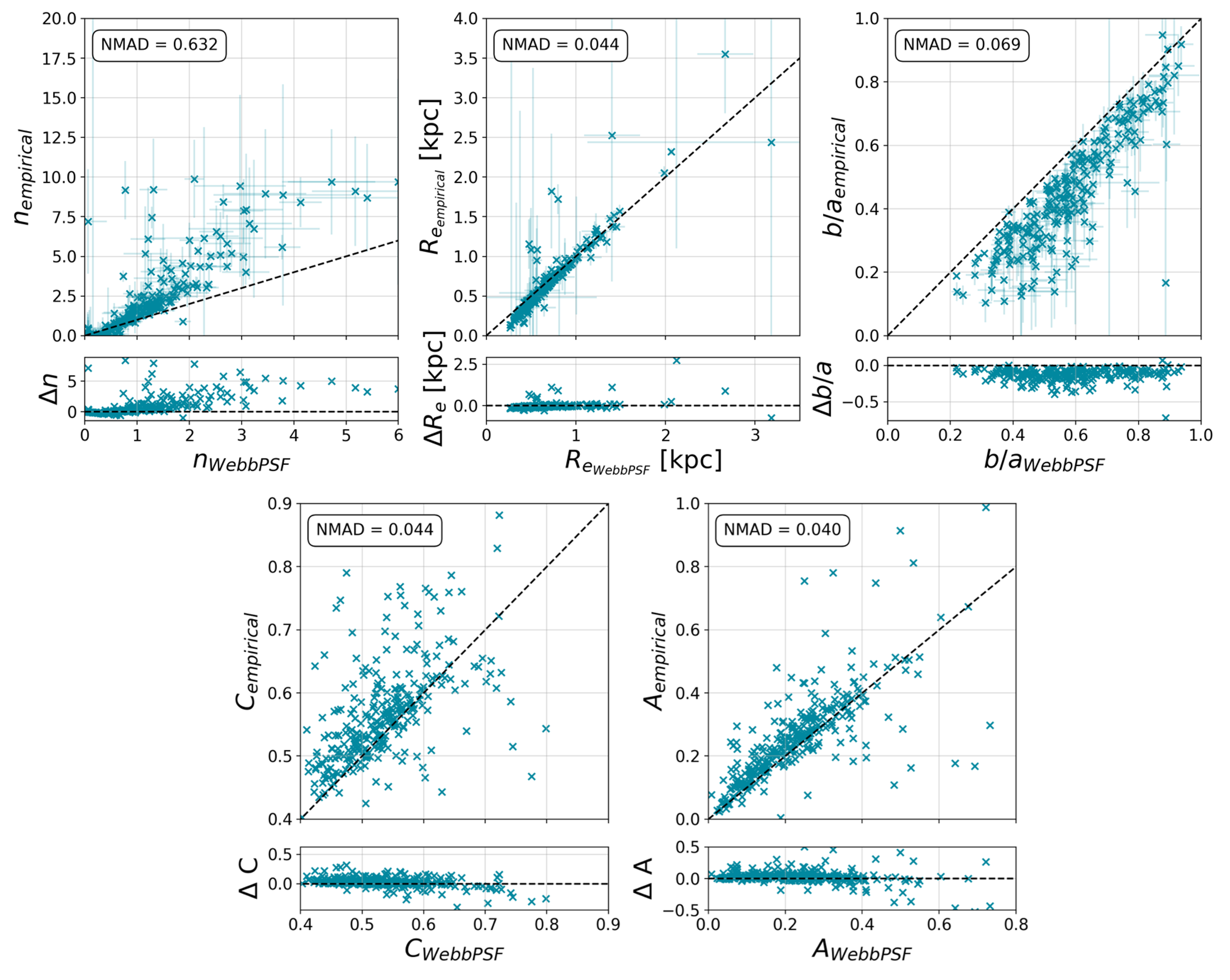}
    \caption{Comparison of measured structural parameters with different PSF models. Teal data points show parameters taken from single S\'ersic models which pass our selection criteria outlined in section \ref{sec:GALFIT criteria}. The bottom panel in each plot shows the difference between the successfully measured parameters of the two PSF models, with positive deltas indicating an increase when measured using the empirical PSF model. NMAD values, as calculated through equation \ref{eq:NMAD} are shown in the top left corner of each plot.}
    \label{fig:PSF_comparison}
\end{figure}

Within this epoch of the universe, the influence of the PSF becomes more and more important, since galaxy sizes start to become comparable in size to the PSF. To test the robustness of our measured structural parameters, we make a comparison of these parameters using the simulated \texttt{WebbPSF} model and an empirical PSF model created by stacking stars from CEERS pointings. Our comparisons are shown in Fig. \ref{fig:PSF_comparison}.

In this comparison, it is clear that certain parameters are sensitive to the PSF model used at these redshifts. The S\'ersic index, in particular, shows a large scatter at higher values of \textit{n} and is the main reason models fail our selection criteria, changing with an average $\Delta n \sim 0.99$ between different PSF models. In addition to this, our measurements of the axis ratio tend to be smaller (less round) by $\Delta b/a \sim 0.1$ when measured with the empirical PSF model, with a low NMAD value.

Despite this, our measurements for $R_{e}$ are in good agreement between the PSF models, with an average difference of $\Delta R_e \sim -0.01$ kpc, showing that these size measurements are less sensitive to the PSF used and robust for this analysis.

We further find there to be little difference on average between non-parametric parameters, with concentration and asymmetry both deviating from the WebbPSF model by $\sim 0.03$ on average.

\section{Potential Mass Break}
\label{sec:mass-break}

\begin{figure}
    \centering
    \includegraphics[width=1\linewidth]{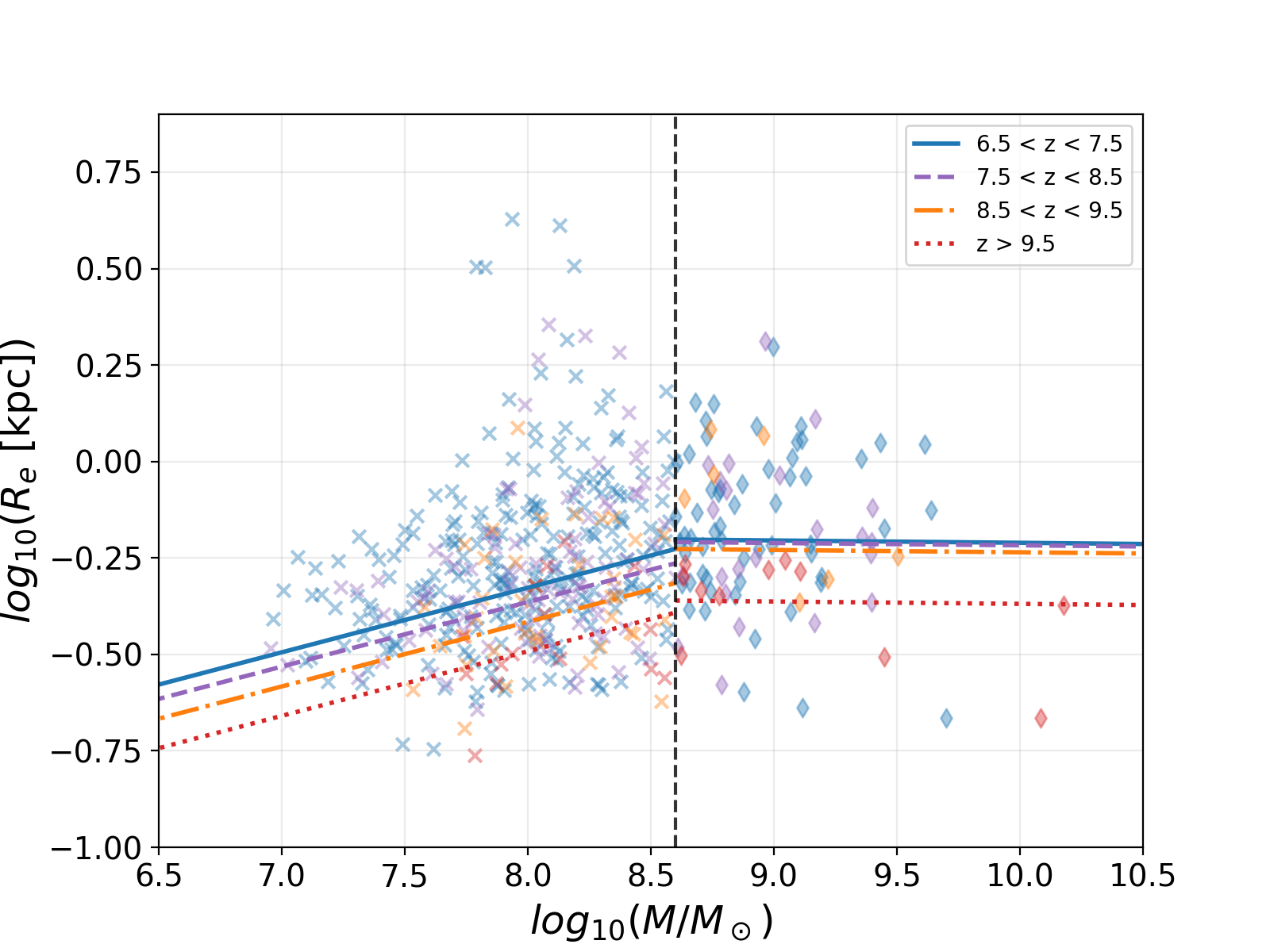}
    \caption{Size-mass relation, split into two mass bins as defined in \citet{chamba_2024}, with best fit lines for each redshift bin. For each bin, the gradient is fixed based on the best fitting line for the lowest Redshift bin of $6.5 < z < 7.5$. Data points are individual galaxies, coloured by their corresponding redshift bin, following the same colour scheme as the best fitting lines.}
    \label{fig:size-mass_mass_break}
\end{figure}

\begin{figure}
    \centering
    \includegraphics[width=1\linewidth]{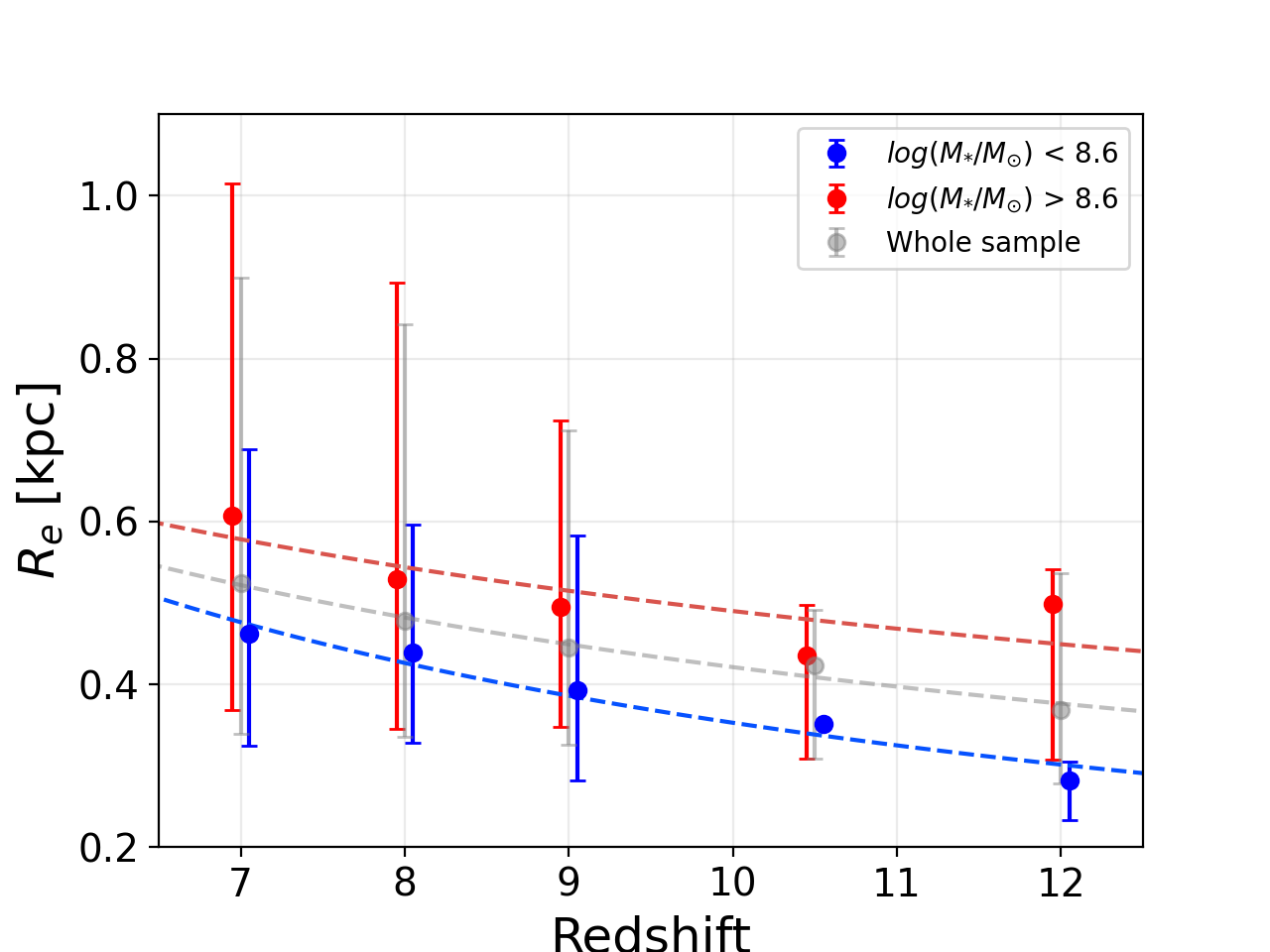}
    \caption{Half-light radii redshift evolution of sample split into two mass bins, separated at $log(M_*/M_{\odot}) = 8.6$. Red data points indicate objects above this mass break, while blue data is related to those below the mass break. Grey data correlates to the whole sample.}
    \label{fig:re-z_mass_break}
\end{figure}

As discussed in \S \ref{sec:results}, it has been reported that galaxies which fall below a mass of $4 \times 10^8 M_{\odot}$ may experience more rapid size growth due to feedback and environmental processes \citep{chamba_2024}. We test this by splitting our sample into two mass bins, separated at this reported break. We show our size-mass relation for these mass bins in Fig. \ref{fig:size-mass_mass_break} and our size evolution in Fig. \ref{fig:re-z_mass_break}. We find a similar steeper gradient in our size-mass relation for lower mass objects, suggesting lower mass galaxies build up their sizes more rapidly than their more massive counterparts. We further observe more massive galaxies to have larger sizes, as expected, however these larger sizes still fall below extrapolated power laws of lower redshift observations, so do not alter our results significantly.

\section{Simulated S\'ersic profiles}
\label{sec: Simulated Sersic Profiles}
In order to test how reliably \texttt{GALFIT} is able to model the surface brightness profiles of objects which are comparable in size to the PSF, we create 4000 single S\'ersic surface profile models with known parameters, using the \texttt{astropy} Python package, which we then convolve with the F444W PSF and add noise to simulate the profile as if being imaged with NIRCam. We then measure this simulated image with \texttt{GALFIT} to see how reliably we recover the structural properties of our models.

Our single S\'ersic surface brightness models consist of a random distribution of structural parameters, ranging between: $0 < n < 5$; $0 < R_{e} ~$[pixels]$~ < 20$ and $0 < b/a < 1$. These are distributed at random for 1000 models at 4 different S\'ersic amplitude values (0.1, 0.05, 0.01 and 0.005) chosen to emulate the SNR levels seen in our data, giving us 4000 models in total at differing levels of SNR.

\begin{figure}
    \centering
    \includegraphics[width=\linewidth]{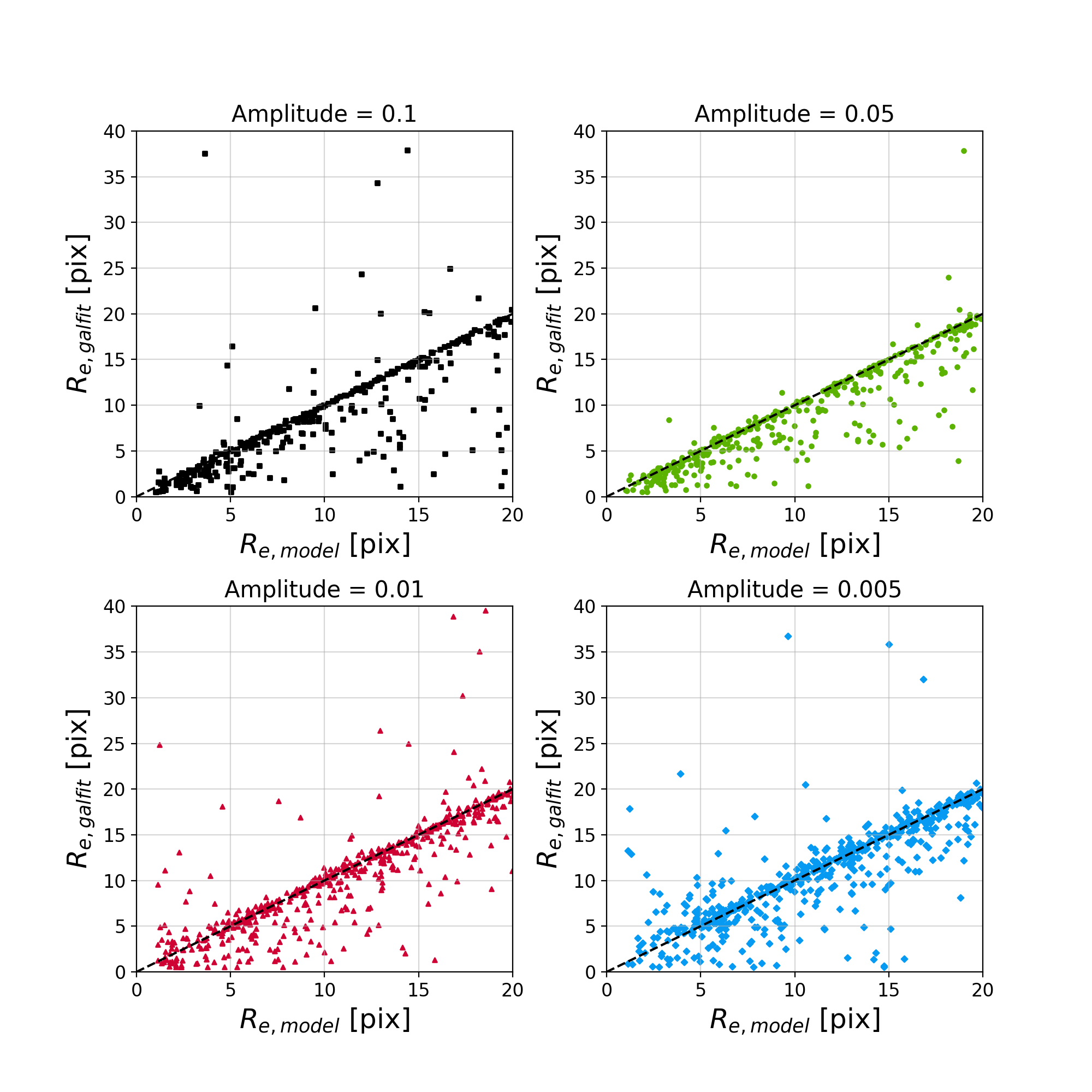}
    \caption{Plot comparing the intrinsic half light radii values of known S\'ersic surface brightness profiles, created through \texttt{Astropy}, and \texttt{GALFIT} measured values of $R_{e}$ at different S\'ersic model amplitudes. Black squares correspond to S\'ersic models of amplitude 0.1; green circles of amplitudes 0.05; red triangles to amplitudes of 0.01 and blue diamonds to amplitudes of 0.005.}
    \label{fig:re_models}
\end{figure}

We find that for all amplitudes of our S\`ersic models, we are able to successfully recover $R_{e}$ of these models in most cases, at all radii, as seen in Fig. \ref{fig:re_models}. In those cases where we do not recover correctly, we find that we underestimate $R_{e}$, in some cases by more than 50\%. Despite this, we find good agreement amongst the majority of the sample.

\begin{figure}
    \centering
    \includegraphics[width=\linewidth]{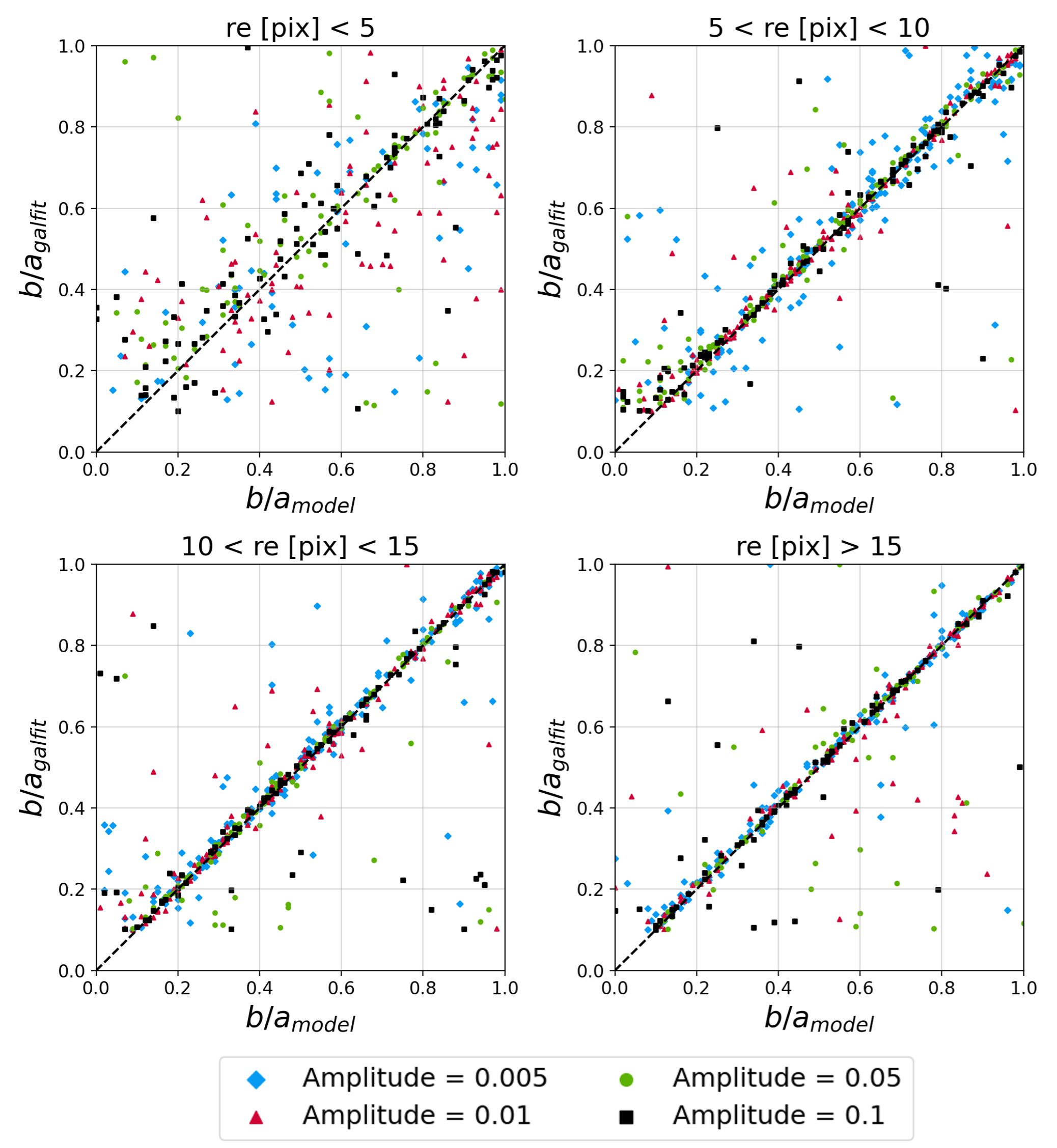}
    \caption{Plot comparing the intrinsic axis ratio values of known S\'ersic surface brightness profiles, created through \texttt{Astropy}, and \texttt{GALFIT} measured values of $b/a$ at different S\'ersic model amplitudes. Black squares correspond to S\'ersic models of amplitude 0.1; green circles of amplitudes 0.05; red triangles to amplitudes of 0.01 and blue diamonds to amplitudes of 0.005.}
    \label{fig:ar_models}
\end{figure}

Further to this, Fig. \ref{fig:ar_models} shows we recover the axis ratios of our models well, in particular when our models have $R_{e} > 5$ pixels. We find slightly worse fits for smaller models, where axis ratio values can change significantly if the radial components of the \texttt{GALFIT} model are slightly off the intrinsic model by a few pixels. This could explain why we find rounder objects in section \ref{sec:results}, as these compact objects are typically on the scale of $R_{e} < 10$ pixels.

\begin{figure}
    \centering
    \includegraphics[width=\linewidth]{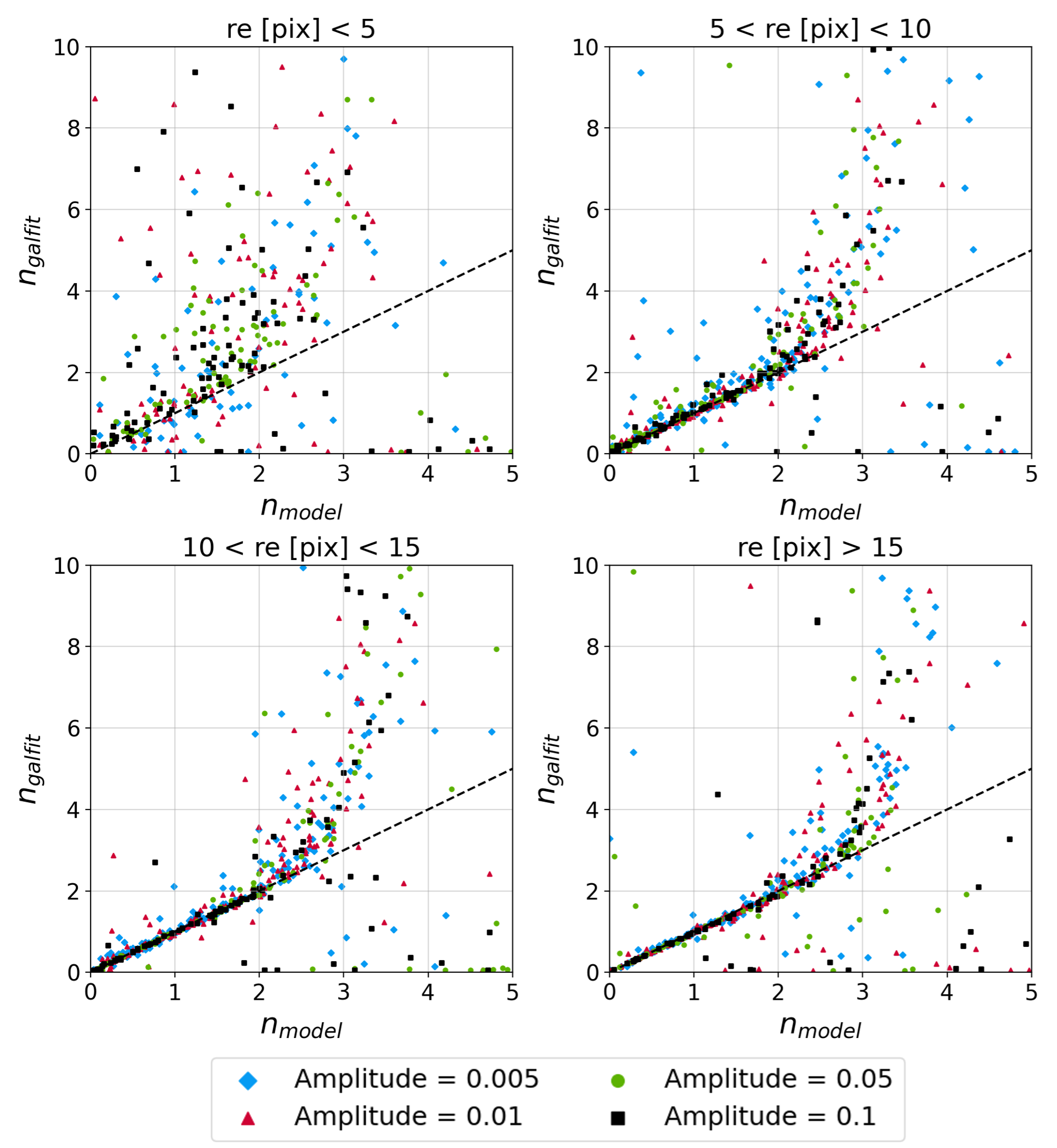}
    \caption{Plot comparing the intrinsic S\'ersic index values of known S\'ersic surface brightness profiles, created through \texttt{Astropy}, and \texttt{GALFIT} measured values of $n$ at different S\'ersic model amplitudes. Black squares correspond to S\'ersic models of amplitude 0.1; green circles of amplitudes 0.05; red triangles to amplitudes of 0.01 and blue diamonds to amplitudes of 0.005.}
    \label{fig:n_models}
\end{figure}

Although we can reliably recover the values of $R_{e}$ and $b/a$ for our models, Fig. \ref{fig:n_models} shows that values of the S\'ersic index of our models on these size scales are not recovered as successfully, particularly for models with higher values of $n$ ($n \geq 3$). At these values, our measured values of $n$ become much larger, hit the lower limit of $n = 0.05$ or fail completely. This could explain why we see a large scatter within our reported distribution of $n$ in section \ref{sec:re evolution}.

In general, we find that we can reliably recover the values of $R_{e}$ for models and objects of all sizes, and the values of $b/a$ for objects.  

\end{document}